\documentclass[prx,aps,twocolumn,superscriptaddress,longbibliography]{revtex4-1}
\usepackage[pdftex]{graphicx}
\usepackage{dcolumn}
\usepackage{bm}
\usepackage{epsfig}
\usepackage{latexsym}
\usepackage{amsmath}
\usepackage{amssymb}
\usepackage{color}
\usepackage{array}
\usepackage{bbm}
\usepackage{hyperref}
\usepackage{color}
\usepackage{array}
\usepackage{cancel}
\usepackage{ulem}
\usepackage{slashed}

\usepackage{hyperref}
\hypersetup{
    linktocpage, colorlinks, allcolors=blue
}

\usepackage{booktabs}
\newcolumntype{C}[1]{>{\centering\arraybackslash}m{#1}}
\AtBeginDocument{
\heavyrulewidth=.08em
\lightrulewidth=.05em
\cmidrulewidth=.03em
\belowrulesep=.65ex
\belowbottomsep=0pt
\aboverulesep=.4ex
\abovetopsep=0pt
\cmidrulesep=\doublerulesep
\cmidrulekern=.5em
\defaultaddspace=.5em
}
\usepackage{booktabs}
\newcolumntype{R}[1]{>{\raggedleft\arraybackslash}p{#1}}
\AtBeginDocument{
\heavyrulewidth=.08em
\lightrulewidth=.05em
\cmidrulewidth=.03em
\belowrulesep=.65ex
\belowbottomsep=0pt
\aboverulesep=.4ex
\abovetopsep=0pt
\cmidrulesep=\doublerulesep
\cmidrulekern=.5em
\defaultaddspace=.5em
}

\newcommand{\<}{\langle}
\newcommand{\e}{\varepsilon}
\newcommand{\up}{\uparrow}
\newcommand{\down}{\downarrow}
\renewcommand{\>}{\rangle}
\renewcommand{\(}{\left(}
\renewcommand{\)}{\right)}
\renewcommand{\[}{\left[}
\renewcommand{\]}{\right]}
\renewcommand{\v}[1]{\vec{#1}} 

\newcommand{\bs}[1]{\boldsymbol{#1}}
\renewcommand{\d}{\partial}

\newcommand{\eps}{\epsilon}

\newcommand{\Z}{\mathbb{Z}}
\newcommand{\T}{\mathcal{T}}
\newcommand{\C}{\mathcal{C}}
\renewcommand{\S}{\mathcal{S}}
\newcommand{\AI}{\rm{AI}}
\newcommand{\AII}{\rm{A{II}}}
\definecolor{DGreen}{rgb}{0, 0.7, 0.0}

\newcommand{\be}{\begin{equation}}
\newcommand{\ee}{\end{equation}}
\newcommand{\bea}{\begin{eqnarray}}
\newcommand{\eea}{\end{eqnarray}}
\newcommand{\nn}{\nonumber \\}

\begin{document}
\title{Realizing topological surface states in a lower-dimensional flat band} 

\author{Andrew C. Potter}
\affiliation{Department of Physics, University of Texas at Austin, Austin, TX 78712}

\author{Chong Wang}
\affiliation{Department of Physics, Harvard University, Cambridge, MA 02138, USA}

\author{Max A. Metlitski}
\affiliation{Perimeter Institute for Theoretical Physics, Waterloo, ON N2L 2Y5, Canada}

\author{Ashvin Vishwanath}
\affiliation{Department of Physics, Harvard University, Cambridge, MA 02138, USA}
\affiliation{Department of Physics, University of California, Berkeley, CA 94720, USA}

\begin{abstract}
The anomalous surface states of  symmetry protected topological (SPT) phases  are usually thought to be only possible in conjunction with the higher dimensional topological bulk. However, it has recently been realized that a class of anomalous SPT surface states can be realized in the same dimension if symmetries are allowed to act in a nonlocal fashion. An example is the particle-hole symmetric half filled Landau level, which effectively realizes the anomalous surface state of a $3d$ chiral Topological Insulator (class AIII). A dual description in terms of Dirac composite fermions has also been discussed.  Here we explore generalizations of these constructions to multicomponent quantum Hall states. Our results include a duality mapping of the bilayer case to composite bosons with Kramers degeneracy and the possibility of a particle hole symmetric integer quantum Hall state when the number of components is a multiple of eight.  Next, we make a further extension by half filling other classes of topological bands and imposing particle hole symmetry. When applied to time-reversal invariant topological insulators we realize a different chiral class (CII) topological surface state. Notably,  half-filling a $3d$ TI band allows for the realization  of the surface of the otherwise inaccessible  $4d$ topological insulator, which supports an anomalous $3d$ Dirac cone. Surface topological orders equivalent to the $3d$ Dirac cone (from the global anomaly standpoint) are constructed and connections to Witten's SU(2) anomaly are made. These observations may also be useful for numerical simulations of topological surface states and of Dirac fermions without fermion doubling. 
\end{abstract}
\maketitle


\section{Introduction}
The $d$ dimensional surface of a $d+1$ dimensional symmetry protected topological phase (SPT), exhibits an anomalous implementation of symmetry that is not allowed in purely $d$ dimensional local system with the same symmetry~\footnote{Note: throughout, $d$ will refer to the number of spatial dimensions, which is one-less than the number of space-time dimensions in a relativistic topological field theory description.}. As an example, the surface of a $3d$ topological insulator hosts a single Dirac cone, which cannot occur in a local $2d$ system without breaking time-reversal ($\T$), or particle-hole symmetry ($\S$) -- a well known example of a fermion doubling theorem. 

A previously unnoticed loophole to this rule was recently discovered in the context of the half-filled Landau level (LL).\cite{Son15,Wang15,Metlitski15,Wang16} Namely, in the restricted Hilbert space of the lowest Landau level of $2d$ electrons, with one electron for every two lowest LL orbitals (half-filling), there is a symmetry between describing many-body states as half-filled with particles or half-empty (half-filled with holes). However, unlike ordinary symmetries, which act locally, the particle-hole symmetry (PHS) is defined only in the restricted sub-space of the lowest LL. Since the LL has a non-zero Chern number, one cannot construct a complete, localized basis of orbitals in the lowest LL. Instead, at least one of the orbitals must be extended across the full sample. Hence, the PHS acting in the lowest LL is inherently non-local, and as a result, can evade certain limitations of local symmetries\cite{NonOnsite}. The strange, non-local nature of the PHS can be alleviated by formally embedding the half-filled LL in the zeroth LL of a massless Dirac electron found at the surface of a chiral topological insulator (class AIII\cite{Kitaev09,Ryu10}) where the non-local nature of the symmetry is reproduced by the anomalous properties of the TI bulk, so that the symmetry acts locally on the TI surface states. The theoretical power of this description is enhanced by a dual description of the chiral TI surface in terms of composite fermionic vortices filling a dual Dirac cone\cite{Son15,Wang15,Metlitski15,Wang16,Mross15}, which is related to a variety of other particle-vortex dualities\cite{karch2016particle,seiberg2016duality}. This dual description enables non-perturbative insights into the possible strongly correlated phases of the half-filled LL. 


In this paper, we ask whether other $(d+1)$-spatial-dimensional SPT surface states can be realized in one lower dimension at the expense of having a non-local action of symmetry, by half-filling various types of $d$-dimensional flat-bands with non-trivial topology. Our motivation is two-fold. First, topological flat bands are naturally realized in various physical contexts (e.g. $2d$ electrons in a strong magnetic field), where they exhibit an approximate particle-hole symmetry.
 The perspective of a higher dimensional SPT surface state, and dual descriptions (for special cases) in terms of composite particles in zero field, allows one to use physical insight from traditional approximation schemes, and non-perturbative results from topological quantum field theory developed in the study of SPT surfaces, in order to understand possible phases of the lower dimensional system.
Such partially filled flat-band systems are inherently strongly correlated, and can exhibit intriguing phenomena from fractional quantum Hall effects to high-temperature superconductivity~\cite{iglovikov2014superconducting}. However, given the inherent non-perturbative nature of the interactions in these settings, developing a controlled theory of such correlated flat-bands remains a daunting challenge. The higher-dimensional surface state correspondence we identify provides rare, topologically robust and non-perturbative insights into these complicated highly fluctuating systems.

A second motivation, is that the inter-dimensional correspondence allows one to numerically simulate the anomalous surface of a higher dimensional SPT phase, without the computational overhead of also simulating bulk degrees of freedom -- thereby avoiding certain fermion doubling issues. Circumventing fermion doubling restrictions could be beneficial not only for simulating condensed matter systems, but also as a means to simulating strongly coupled chiral matter, such as QCD systems.
  

\section{General conditions for correspondence between flat-bands and anomalous TI surface states}
We start by formulating a set of general conditions for an inter-dimensional correspondence, in which a non-local PHS of a topological flat band matches the anomalous behavior of the surface state of a TI in one higher dimension. The prototypical example of such an inter-dimensional correspondence, is that between a half-filled Landau level and a chiral TI (class AIII) surface state with a single Dirac cone. To recall, in the limit of large inter-LL gaps one can effectively project into the Hilbert space spanned by the orbitals of the lowest LL. This restricted Hilbert space exhibits a non-local PHS upon half-filling, which precisely matches the Hilbert space and PHS symmetry of the zeroth LL of the chiral TI surface state in a magnetic field (note that for class AIII, the magnetic field does not break the protecting PHS). Our criterion will enable us to identify a number of previously unconsidered examples of topological flat bands that are related to higher-dimensional TI surface states.

The half filled Landau level example utilize free fermion topological band structures, with nontrivial Chern number (class A). Imposing particle hole symmetry by projecting to a single, half filled band and considering only two body interactions (i.e. flat dispersion) mapped this problem to the single-Dirac cone surface state of a $3d$ class AIII topological insulator. The anomalous realization of particle hole symmetry, as a nonlocal symmetry in the projected bands, was key to the success of this procedure. Therefore we will choose to generalize this procedure by considering other topological band structures for fermions that we will choose to half fill, to obtain an anomalous particle hole symmetry.  The necessary ingredients then are:
\begin{enumerate}
\item Without interactions, the system must posses a non-trivial topological band structure for the free fermions.
\item The system must possess a U(1) charge conservation symmetry to impose a fixed filling.
\item The topological nature of this band structure must not rely on a local particle-hole symmetry, since we will be placing the chemical potential in the center of the band rather than at the natural $E=0$ (though an additional local particle-hole symmetry may be present at half-filling, as happens for graphene, it must not play a defining role in the band topology).
\end{enumerate}
We will further restrict our attention to physically accessible number of spatial dimensions $d=1,2,3$
\footnote{We exclude $0d$ cases, since there is no suitable distinction between local and ``non-local" symmetries in this case.}.
Half filling a flat topological band satisfying these conditions imposes an additional particle hole symmetry $\S$, which is an anti unitary operator (when acting on many body states), defined by: $\S c^\dagger_{r\sigma}\S^{-1}=c^{\vphantom\dagger}_{r\sigma}$, where $c^\dagger_{r,\sigma}$ creates an electron with spin/orbital/flavor label $\sigma$ at position $r$. We can verify that the antiunitary nature of $\S$ is required by considering the transformation properties of the creation operator of a generic single particle orbital, $n$, $\psi_n = \sum_{r\sigma} \phi^n_{r\sigma} c_{r\sigma}$, on which $\S$ acts as:
\begin{align}
\S \psi_{n}^{\vphantom\dagger}\S^{-1}&= \S \sum_{r\sigma} \phi^n_{r\sigma} c^{\vphantom\dagger}_{r\sigma} \S^{-1}
= \sum_{r\sigma} \phi^{n*}_{r\sigma} c^\dagger_{r\sigma} =\psi_n^\dagger
\end{align}
i.e. the anti-unitary operator $\S$ exchanges generic orbital creation and annihilation operators, and hence commutes with the Hamiltonian of the half-filled flat band.

Given (1), the starting point is a free fermion band structure. For simplicity, we will restrict our attention to the 10 fold way classification of the subset of non-interacting fermion phases that incorporate only time reversal $\T$ and charge conjugation $\C$ symmetries, and will neglect the role of other on-site or crystalline symmetries.

\begin{table}[htp]
\begin{center}
\begin{tabular}{|c||cccccccc|}
\hline 
Symmetry Class & & & & dim & & & &\\
&0&1&2&3&4&5&6&7\\
\hline
 A & Z & 0 & {\color{DGreen} \bf Z} & 0 & Z &0&Z &0\\
 AIII & 0 & Z & 0 & {\color{DGreen} \bf Z} & 0 & Z &0&Z \\
 \hline\hline
  AII & 2Z & 0 & {\color{red} {\bf Z}$_2$} &{\color{blue} \bf Z$_2$} & Z &0&0 &0\\
 CII & 0 & Z & 0 & {\color{red} \bf Z$_2$} & {\color{blue} \bf Z$_2$} & Z &0&0 \\
 \hline 
 \hline
   AI & Z & 0 & 0 &0 & 2Z &0&Z$_2$  &Z$_2$\\
 BDI & Z$_2$  & Z & 0 & 0 & 0 & 2Z &0&Z$_2$ \\
 \hline
 \end{tabular}
\end{center}
\caption{Noninteracting fermion topological classification in various dimensions for the relevant symmetry classes (A, AII and  AI) and the ones connected to them by adding a chiral symmetry $\S$, classes (AIII, CII and BDI respectively). Half filling a topological flat band in the first set realizes the surface state of the corresponding topological phase in the second class, in one higher dimension. The cases studied in paper are indicated in bold face.  In this section we focus on flat bands of the class AII topological insulators, in $2d$ and $3d$ (shown in red and blue). They map to the surface of topological phases of class CII in $3d$ and $4d$. Note: all dimensions refer to the number of spatial dimensions, i.e. are one less than the number of spacetime dimensions.}
\label{TableKtheory}
\end{table}

Incorporating points (2,3) above leaves us with just three candidate symmetry classes $\rm{A}$, $\AI$ and $\AII$ for the $d$ dimensional flat-band system. Crucially, due to the shifting structure of phases in the free fermion 10-fold way classification, in each of these cases (see Table~\ref{TableKtheory}), there is a corresponding class in one higher dimension with the same free-fermion classification. Denoting the free-fermion classification in symmetry class $X$ in $d$ dimensions as $C_d(X)$, we summarize the relevant relations:
\begin{align}
C_d(\rm{A}) &= C_{d+1}(\rm{AIII})\nonumber\\
C_d(\rm{AI}) &= C_{d+1}(\rm{BDI})\nonumber\\
C_d(\rm{AII}) &= C_{d+1}(\rm{CII})
\label{eq:correspondence}
\end{align}
These relationships show that, whenever there is a non-trivial $d$-dimensional flat-band with a non-local PHS, then, there is a corresponding non-trivial higher-dimensional TI surface state with a local, but anomalous implementation of that PHS. A second necessary property for realizing the higher dimensional surface state in a flat band is that one should be able to disconnect the surface state from the bulk bands. Of course this can only occur in cases with particle hole symmetry, since otherwise one can simply move the chemical potential to a region without surface states, contradicting the basic assumption that free fermion topological phases must have gapless boundary modes. For previously discussed examples, this disconnection could be done simply by applying a magnetic field to split the surface into Landau levels. For time-reversal invariant classes, we will describe an alternative scheme for surface state detachment in Appendix~\ref{app:disconnect}.

With these ingredients in place, we are now in position to explore the correspondence suggested by Eq.~\ref{eq:correspondence}. Class AI has non-trivial topological bands only high dimensions $d~\text{mod}~8 =4,6,7,8$, and can be ignored if we are interested in physical systems ($d\leq 3$). In each of the remaining cases for A and AII bands, we will find an inter-dimensional correspondence. In the main text, we provide a detailed and pedagogical exploration of this correspondence and its implications for half-filled flat bands. Before doing so, we briefly summarize the main results.

\subsection{Half filled Chern bands (A) and chiral TIs (AIII)}
Class A has non-trivial topological bands in $2d$, classified by an integer valued Chern number, and potentially corresponds to the surface state of a $3d$ AIII. This surface state is characterized by an integer chirality, $N$ that represents the number of chiral surface Dirac cones. We have already encounter an example of the correspondence, Eq.~\ref{eq:correspondence} in this case: the half-filled LL (Chern number one), which corresponds to the single-Dirac cone AIII surface state. By analogous reasoning, one can readily verify that these correspondence between a $2d$ half-filled Chern number-$N$ and the chirality-$N$ $3d$ surface state extends to all $N$. This correspondence enables us to leverage known resuls about the properties of interacting AIII surface states[REFs] to gain insights into the behavior of seemingly unrelated multicomponent quantum Hall systems such as multilayer graphene at neutrality, or multi-layer GaAs quantum wells in a strong field.

\begin{figure}[t!]
\includegraphics[width=\columnwidth]{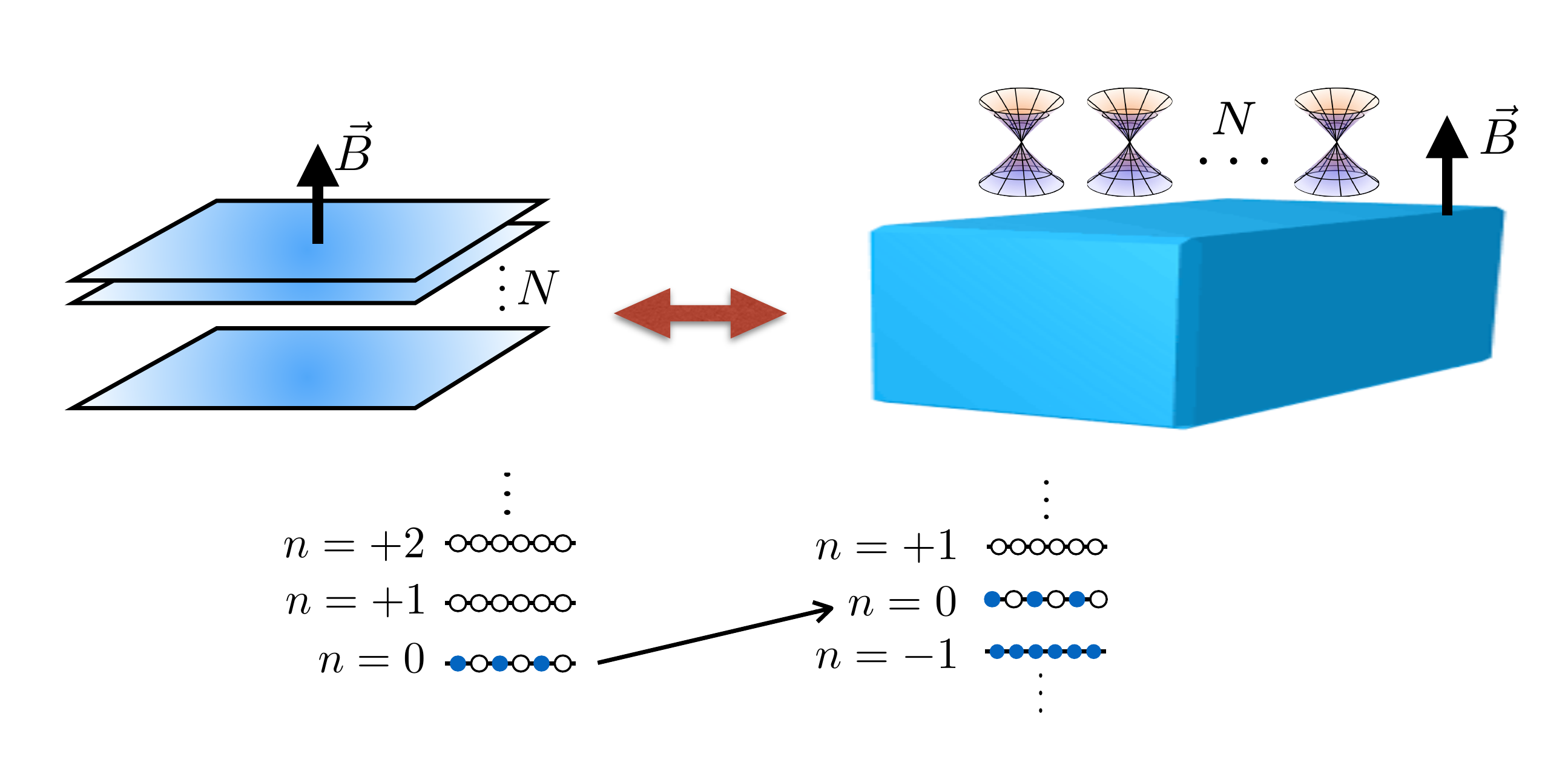}\vspace{-.1in}
\caption{ {\bf Half filled flat band/TI surface state correspondence -- } N$\times$ half-filled Landau levels of $2d$ electrons, corresponds to the zeroth Landau levels of the N-Dirac cone surface state of a $3d$ chiral (class AIII) topological insulator. }
\label{fig:HigherD}
\vspace{-.1in}
\end{figure}

Interestingly, while the free-fermion band-classifications match, the interacting classifications do not. Whereas the Chern number remains robust to interactions, the chiral TI surface state collapses modulo $N=8$ in interacting systems. We show that this enables an unconventional correlated integer quantum Hall state of 8-half filled LLs (e.g. bilayer graphene at neutrality, and neglecting spin-splitting), which preserves the non-local PHS.

To demonstrate this correspondence, we will begin by examining the half-filled lowest LLs of bilayer quantum Hall systems, such as spin polarized graphene in a magnetic field at neutrality, or GaAs double wells at half-filling in a strong magnetic field. We show that these half-filled bilayers realize the surface state of a $3d$ chiral TI  with two surface Dirac cones (protected by $\S$ symmetry, i.e. in class AIII). Unlike the helical time-reversal symmetry protected TI (class AII) familiar from Bi$_2$Se$_3$ and related materials\cite{HasanKaneRMP} for which an even number of surface Dirac cones would be topologically trivial, the double-Dirac cone surface state of the chiral TI is anomalous and cannot be realized in a local $2d$ system. We further show that the half-filled bilayer permits a dual description in terms of composite bosons, and analyze possible correlated phases of this system from this dual perspective. A related duality between bosonic SPT surface states and 2-component QED was previously explored~\cite{Senthil04,Bi15,mross2016bosonic}. While conceptually similar, this description does not directly connect to the lower-dimensional electronic flat-band system, which corresponds to an \emph{ungauged} $N=2$ Dirac cone system, or a \emph{gauged} boson SPT surface state. We will see that this distinction changes the nature of symmetry preserving fractional quantum Hall states that are possible for the half-filled $N=2$ flat-bands.
From the dual composite boson perspective, we show that it is possible for the half-filled bilayer to preserve the non-local $\S$ symmetry by forming an Abelian quantum Hall state with $\Z_4$ topological order, which has an anomalous realization of $\S$ symmetry.  This anomalous surface topological order was previously obtained directly for the interacting two-Dirac cone surface of the chiral Fermion (class AIII) TI surface~\cite{Wang13Window}. Here, we reproduce these results from the dual composite boson perspective to demonstrate equivalence of these two dual descriptions.

Next we generalize this construction to arbitrary number of layers. For example, the half-filled quadruple layer quantum-Hall system (or bilayer graphene in a strong field at charge neutrality), corresponds to a non-trivial chiral TI surface with four Dirac cones, from which we may identify new candidate phases including an Abelian quantum Hall state with $\Z_2$ topological order, and a gapless phase when enriched by an enlarged $SO(5)\times \S$ symmetry\cite{Senthil04,LeeSachdev14,Wu14,LeeSachdev15,Nahum15}. 
 We further find that strong interactions can drive the half-filled octuple layer quantum-Hall system into a non-fractionalized integer quantum Hall state, without breaking the non-local $\S$ symmetry.

\subsection{Half-filled helical TI bands (AII) and heli-chiral TIs (CII)}
Besides the correspondence between half-filled Chern bands and Chiral TI surface states, our general criterion predicts a second, previously unrealized correspondence between flat time-reversal symmetric (``helical", class AII) flat-bands and a higher dimensional heli-chiral TI surface states (class CII). 

Class AII has non-trivial band topologies in $2d$ and $3d$, exemplified by materials such as HgTe and Bi$_2$Se$_3$ respectively, and characterized by well known $\Z_2$ topological band invariant~\cite{HasanKaneRMP}. Though these AII TIs typically arise in weakly correlated materials, there are a growing number of materials with strong spin-orbit coupling and strong correlations that may realize these phases. In systems like SmB$_6$, topological insulation is theorized to arise from hybridization between conduction electrons and local f-moments. This hybridization can result in very narrow ``heavy fermion" bands, for which interactions can greatly exceed band-width. Half filling such a half-filled heavy fermion band would result in an approximate non-local PHS, and which can lead to exotic fractional phases, that can be described from the higher-dimensional TI surface state perspective. Electrons in class $\AII$ have conserved charge $N_F$, and time reversal symmetry with $\T^2=(-1)^{N_F}$. Combining the non-local PHS, $\S$, with time reversal yields a unitary (in the many-body sense) particle-hole symmetry: $\C=\S \T$, which satisfies $\C^2 = \T^2=(-1)^{N_F}$. Therefore the resulting symmetry class has $\C^2=\T^2=-1$ which corresponds to the symmetry class CII. 

We show that $2d$ and $3d$ half-filled AII flat bands, indeed correspond to the anomalous CII surface states in $3d$ and $4d$ respectively. To do so, we show that the surface states of the CII TI can indeed by disconnected from the bulk states, and flattened to achieve a one-to-one correspondence with the lower-dimensional flat band. One additional subtlety is that, unlike the half-filled Landau level system, for which any two-body interactions preserve the non-local PHS, for this time-reversal symmetric (TRS) class, only certain types of interactions preserve the PHS. We derive a general criterion, and show that the presence of common crystalline symmetries can ensure that PHS is maintained in the interacting system.

For the $2d$ AII flat band, we carefully re-examine the physics of the corresponding $3d$ CII TI, and correct previous incorrect claims from the literature[REF]. This updated understanding of the CII TI enables us to derive a dual description of the $2d$ flat band in terms of charge-neutral composite boson vortices that transform anomalously as $\C^2=-1$, and explore possible fractionalized topologically ordered states from this dual perspective.

Notably, the same correspondences works equally well in one higher dimension. Namely, a $3d$ AII flat band enables the realization of an anomalous $3d$ surface state of a $4d$ CII TI. This opens the door to the study of anomalous $3d$ topological orders, which had previously not had a physical context in our $3d$ world. We show that this $3d$ system with a non-local PHS realizes a close relative of Witten's well-known $SU(2)$ anomaly\cite{wittenanomaly}, and construct a topologically ordered state with a fractional axion angle that preserves the anomalous PHS of this $3d$ system.

\section{Half-filled  multilayer quantum Hall systems}
We begin by examining N-layer $2d$ electrons with non-relativistic dispersion in a external magnetic field that is sufficiently strong in order that the inter-LL gaps greatly exceed the interaction energy scale, and with particle density equal to half of the total number of orbitals in the lowest-Landau levels of the multilayer system (half-filling). In the limit of very large B-field, we may restrict ourselves to the orbitals of the lowest LL, in which case the $2d$ multilayer lowest LL Hilbert space is identical to that arising in the zeroth-Landau levels of N-Dirac fermions at the surface of a chiral TI (symmetry $U(1)\times \S$, class AIII) in a strong magnetic field (Fig.~\ref{fig:HigherD}). 

In the absence of a magnetic field, the surface Dirac cones of the corresponding N-component class AIII TI have a Hamiltonian of the form:
\begin{align}
H_\text{AIII} = \sum_{a=1}^N \psi^\dagger_a \(p_x\sigma^x+p_y\sigma^z\)\psi^{\vphantom\dagger}_a
\end{align}
where $a$ is a flavor label ranging from $1$ to $N$, each $\psi_a$ is a two-component $2d$ Dirac fermion, $\boldsymbol{\sigma}$ are Pauli matrices operating in the psuedospin space of each Dirac flavor. This surface state is protected by an anti-unitary particle hole symmetry, $\S=\C\T$ which can be viewed as the combination of time-reversal, $\T$ and unitary particle hole conjugation $\C$. Namely, the particle-hole symmetry acts like:
\begin{align}
\S\psi_a = i\sigma^y K\psi_a^\dagger
\label{eq:CT}
\end{align}
where $K$ indicates global complex conjugation, which forbids all fermion bilinear mass terms: $\sum_{a,b}\psi^\dagger_a \sigma^ym_{ab}\psi_b$.

An external magnetic field, described by substituting $\v{p}\rightarrow \v{p}-\v{A}$ with $\v{\nabla}\times \v{A} = B(\v{r})\hat{z}$, preserves the $\S=\C\T$ symmetry, since $B$ is flipped by each of $\C$ and $\T$. Hence, the full spectrum of the Dirac Landau levels of the AIII surface is $\S$ symmetric. However, in the corresponding non-relativistic $2d$ system, only the lowest LL obeys $\S$ symmetry. Since the orbitals of the lowest LL are not all localized, the symmetry that interchanges these orbitals being occupied or unoccupied is similarly non-local. As explained above, this non-locality enables the purely $2d$ system to realize the physics of an anomalous higher dimensional TI.

A second beneficial aspect of this relation is that, formulating a manifestly $\S$ invariant description of the purely $2d$ flat-band requires working within a non-locally constrained Hilbert space (projected into the lowest Landau level), presenting daunting theoretical challenges. While it may be possible to construct an explicitly lowest LL projected theory of the half-filled LL of fermions (e.g. following related approaches for bosonic particles at filling fraction one\cite{Pasquier98,Read98}), successfully implementing such a program remains an outstanding challenge. Moreover, even if successful, the analysis of such an explicitly lowest LL projected description is likely to be complex and involved\cite{Read98}, and extracting the physical properties of such a state can be challenging (however, see Refs.~\cite{ShankarMurthy,WS06} for recent attempts). Instead, viewing the half-filled LL from the perspective of a higher-dimensional TI surface, in which $\S$ symmetry is locally implemented, allows one to use familiar field theoretic tools and apply substantial physical intuition built from ones understanding of simpler systems, to identify possible phases for the half-filled LL and compute their properties in a manifestly $\S$ invariant manner. This latter advantage is substantially enhanced by the discovery of dual descriptions that allow one to trade strong magnetic field for finite density of dual vortex objects\cite{Son15,Wang15,Metlitski15,Wang16,Mross15}  -- allowing dual description in terms of particles in zero net field, skirting the thorny problem of contending with the non-commuting geometric structure of the lowest LL.

\subsection{Half-filled single layer $(N=1)$}
We begin by reviewing the single-component case, which was previously studied in Refs.~\cite{Son15,Wang15,Metlitski15,Wang16,Mross15}. A single-component Landau level at half filling, in the strong field limit such that inter LL mixing is negligible, corresponds precisely to the zeroth Landau level of the single-Dirac cone surface state of a chiral TI in a magnetic field. Since the Hilbert spaces of these two problems coincide precisely, on the face of it, analyzing either partially filled dispersionless band is equally hard. However, the $3d$ TI surface theory permits a non-perturbative dual description in terms of neutral composite fermion vortices. From this dual perspective, one can identify new natural candidate phases that manifestly preserve the non-local $\S$ symmetry of the half-filled LL, which are difficult to obtain directly in the original electron language, and hence had previously escaped attention.

\subsubsection{Dual composite fermion description}
We briefly review the dual description of the AIII surface state in terms of $4\pi$ flux carrying fermionic vortices. This duality is explained in detail in Refs.~\cite{Son15,Wang15,Metlitski15,Wang16,Mross15}. One way to derive the dual description is to first consider coupling the electrons in the bulk of the $3d$ chiral TI to a compact fluctuating electromagnetic gauge field, $A^\mu(\v{r},t)$\cite{Metlitski15,Wang16U1SL}. The electromagnetic duality of the resulting bulk $U(1)$ gauge theory has, via the bulk-boundary correspondence of topological phases, a corresponding surface duality between charge and flux-like excitations of the surface.

Namely, integrating out the gapped bulk electron excitations of the chiral TI produces a magneto-electric term at $\theta=\pi$ in the bulk action for the gauge field, which (written in Euclidean space-time) reads:
\begin{align}
\mathcal{L}_\text{bulk} = i\frac{\theta}{4\pi^2}\v{E}\cdot\v{B}+\dots
\end{align}
Consequently, the unit-strength monopoles (i.e. with $\int_S \v{B}\cdot d\v{a} = 2\pi$ for any surface, $S$, surrounding the monopole), accumulate half of an electrons worth of polarization charge inside the AIII TI, i.e are dyons with electric and magnetic charge $(q_e,q_m) = (\pm \frac{1}{2},1)$. Since the polarization charge does not alter the exchange statistics, these $(1/2,1)$ dyons are bosonic particles. The minimal electrically neutral magnetic excitation can be obtained by binding two $(1/2,1)$ dyons to form a $(1,2)$ bosonic dyon, and then binding an electron to this composite particle to form a neutral $(0,2)$ double-monopole. The act of binding the extra fermion, required to neutralize the polarization charge from $\theta=\pi$, alters the statistics of the $(0,2)$ particle to be fermionic. 

This magnetically charged fermion is a gapped excitation that is in a non-trivial {\it helical} (class AII) topological insulator band (the same class as the usual $3d$ electronic topological insulator realized in Bi$_2$Se$_3$). To see this, first note that while $\S$ symmetry flips the electric charge, it preserves the magnetic charge. Hence, the symmetry group for the dual $(0,2)$ fermions is $U(1)_m\rtimes \S$ (where $U(1)_m$ refers to the magnetic charge conservation of the $U(1)$ gauge theory), i.e. $\S$ acts like ordinary time-reversal on the dual magnetic fermions, so that the dual fermions are in class AII. To see that the dual fermions form a non-trivial AII state, note that the $(0,2)$ fermion with magnetic charge $q_m=2$, sees a $q_e$-electric charge as a point-like source of $4\pi q_e$ flux. I.e. a half-electric charge, $q_e=1/2$ is seen as a strength-one ``monopole" for the $q_m=2$ fermion. Since the $q_e=1/2$ excitations have unit magnetic charge (i.e. half of the magnetic charge of the dual fermion with $q_m=2$), the $q_m=2$ fermions also exhibit a topological magneto-electric effect corresponding at effective $\theta=\pi$, the hallmark of a topological insulator state.\cite{HasanKaneRMP} Moreover, since the AII topological insulator requires that the dual fermions transform like Kramers doublets under $\S$ symmetry, the $(0,2)$ fermionic double monopole has $\S^2=-1$ (see Ref.~\cite{Wang14Science,Metlitski15STO} for more details). This bulk duality has been more formally justified in Ref.~\cite{MaxSdual} by studying the partition function of the theory on non-trivial manifolds.

Due to the surface-bulk correspondence of topological phases, the bulk electric and magnetic excitations of this $U(1)$ gauge theory also have corresponding surface excitations with the same electric and magnetic charge and statistics, which can be created by dragging the bulk excitation across the surface into surrounding vacuum. One can formally confine the bulk monopoles to go back to the un-gauged AIII topological insulator without confining their surface avatars\cite{Metlitski15}. These thought experiments reveal that the surface state of the AIII topological insulator permits a dual description in terms of neutral $4\pi$-vortex-like fermion excitations (corresponding to the bulk $(0,2)$ dual fermion of the now-confined bulk gauge theory) that couple to a fluctuating $U(1)$ gauge field living only within the surface. Just as for the $(0,2)$ double-monopoles of the bulk gauge theory, $\S$ acts on the dual surface fermions as ordinary time-reversal. Such a dual fermion surface theory is described by a simple Lagrangian
\begin{align}
\mathcal{L}_\text{cf} = \bar{\psi}_\text{cf}(i\slashed{\partial}+\slashed{a}-\mu\gamma_0)\psi_\text{cf}-\frac{1}{4\pi}adA, \label{eq:Lcf}
\end{align}
where the dual Dirac composite fermion $\psi_\text{cf}$ couples to a dynamical gauge field $a_{\mu}$, with a chemical potential $\mu$ adjusted so that the composite fermion density\footnote{Strictly speaking the chemical potential term is not needed (or even meaningful) for a dynamical gauge theory, since the total gauge charge is always zero by gauge invariance. The term is introduced here to enable a mean field description, in which the gauge constraint is satisfied at the mean-field level."} equals to $B/4\pi$, and we have introduced the short-hand notation $adA \equiv \epsilon^{\mu\nu\lambda}a_{\mu}\partial_{\nu}A_{\lambda}$. More formal aspects of the duality, both on the surface and in the bulk, are discussed in Ref.~\cite{seiberg2016duality, metlitski2017intrinsic}.

One can also readily verify that all of the anomalous states of the AIII surface can be achieved in the dual picture, such as the $\S$-breaking quantum Hall insulator\cite{Son15,Wang15,Metlitski15,Wang16}. Explicit derivations of the duality have also been obtained in certain spatially anisotropic coupled-wire constructions\cite{Mross15}.

\subsubsection{Candidate phases for the half-filled LL}
From the perspective of this $3d$ chiral TI surface, we can envision various possible phases for the half-filled LL. A conceptually simple possibility is that the interactions could spontaneously generate a $\S$-breaking mass, resulting in a quantum Hall ferromagnet (QHFM) with Hall conductance either $\sigma^{xy}=0,1$ (or, at fixed filling, phase separated domains of each). Domain walls in this QHFM would then carry chiral fermionic modes corresponding to the edge of a $\nu=1$ integer QH state.

Empirically, however, the half-filled lowest LL of GaAs seems to realize a gapless composite Fermi liquid (CFL) phase, which can be roughly viewed as a quantum disordered QHFM in which the gapless domain walls are proliferated.\cite{Vishwanath13,Mross15} Numerical simulations\cite{RezayiHaldane, Geraedts16} suggest that the CFL preserves $\S$-symmetry. A description of the CFL phase can be obtained\cite{Son15,Wang15,Metlitski15,Wang16} through the duality (\ref{eq:Lcf}) reviewed in the previous section: the CFL possesses a Fermi-surface of double-vortices $\psi_{{\rm cf}}$ of the electronic fluid. 
The crucial difference between the composite fermions viewed as  vortices, compared to those originally obtained by flux attachment via a singular gauge transformation\cite{HLR}, is that $\S$ symmetry is manifestly preserved in this description. The resulting physical difference is that the  composite fermions are neutral particles with an electric dipole moment that transforms under $\S$ symmetry as a spin-1/2 degree of freedom does under time-reversal symmetry. In particular, in the presence of $\S$ symmetry, this dipolar psuedospin is polarized perpendicular to the composite-fermion momentum, giving rise to a quantized Berry phase of $\pi$ around the composite Fermi surface. This predicted Berry phase manifests itself in transport experiments via the quantization of electrical Hall conductance \cite{Son15} and in an anomalous Nernst effect\cite{Potter15}.

In addition, the TI surface perspective suggests other possible $\S$-preserving quantum-Hall states with $\sigma^{xy}=1/2$ that might be potentially realized for the right combination of interaction potentials. For example, the single Dirac cone surface state of the $3d$ chiral TI corresponding to the half-filled LLL can exhibit a fully gapped $\S$ preserving state with intrinsic topological order. One such symmetry preserving surface topological order, dubbed the CT-Pfaffian state, has been constructed in lattice models\cite{Fidkowski13}. 
The CT-Pfaffian state has non-Abelian topological order, with Ising (Majorana)-like non-Abelian excitations. However, unlike the Pfaffian state proposed by Moore and Read for the $\nu=5/2$ plateau, this CT-Pfaffian preserves $\S$ symmetry, and in fact the $\S$ symmetry acts in an anomalous manner on the topological excitations. While small-scale numerical simulations suggest that the $\nu=5/2$ $2d$EG with Coulomb interactions prefers to spontaneously break the underlying $\S$ symmetry, it was recently proposed that impurities and other effects could possible stabilize the alternative CT-Pfaffian order in experimental systems.\cite{Zucker16}

\subsection{Half-filled bilayer $(N=2)$}
Having reviewed the previously studied case of a single-component half-filled Landau level (LL), we now turn to half-filled LLs in systems with two flavors of fermions. One physical example is a GaAs bilayer quantum well subjected to a magnetic field such that each layer is at half-filling\cite{Suen92,Eisenstein92,Murphy94}. Another natural two-component quantum Hall system is single layer graphene subjected to a sufficiently strong magnetic field to spin-polarize the electrons, and with chemical potential tuned such that the dominant-spin LLs are half-filled\cite{GrapheneRMP}. 
Projecting into the lowest LLs, this system also has a non-local $\S$ symmetry that interchanges the states with all lowest LL orbitals full and empty respectively. Applying $\S$ changes the Hall conductance from $\S:\sigma^{xy}\rightarrow \(2-\sigma^{xy}\)$ (in units of $\frac{e^2}{h}$) leaving $\sigma^{xy}=1$ invariant. Just as for the $N=1$ case described above, this LL projected Hilbert space with non-local symmetry is precisely the same as the zeroth LLs of the surface of a $3d$ chiral TI (class AIII) with two surface Dirac cones, where, in the $3d$ system, the $\S$ symmetry extends to the full Dirac spectrum and acts locally. 

We will find a dual description in terms of composite boson vortices, that have a Kramers degeneracy enforced by $\S$. 
Before diving into a detailed description of the duality, let us anticipate this result using the following physical arguments. Let the two species of the $N=2$ system be labeled $c,\,d$. Note, we do not  require that the relative species number is conserved, just that we have a particle hole symmetry, $\S$ that transforms the fermion operators $\{c_i,\,d_i\}\overset{\S}{\longleftrightarrow} \{c^\dagger_i,\,d^\dagger_i\}$ and transforms the vacuum  into the fully filled state $|0\rangle\overset{\S}{\longleftrightarrow} \prod_i c^\dagger_i d^\dagger_i\|0\rangle $. Clearly we have as many fermions as flux quanta (half filled Landau levels with N=2) so that we may obtain a description of particles in zero average field by ``binding" a single unit of flux to each electron, resulting in composite bosonic object. Due to the Hall conductance, the extra flux quantum will suck in a unit of charge that cancels the electron charge, resulting in an overall charge neutral composite boson. It remains to show that the composite bosons are Kramers doublets - which implies that whenever there are an odd number of flux quanta, states must appear is doubly degenerate pairs. This is readily seen for the simplest case of a single flux quantum $N_\phi=1$. The two possible states are $|1\rangle  = c^\dagger |0\rangle,\, {\rm and}\, |2\rangle = d^\dagger|0\rangle$. Now under particle hole symmetry we have $c^\dagger |0\rangle \overset{\S}{\longrightarrow} c (c^\dagger d^\dagger|0\rangle)$ and  $d^\dagger |0\rangle \overset{\S}{\longrightarrow} d (c^\dagger d^\dagger|0\rangle)$. This implies $|1\rangle  \overset{\S}{\longrightarrow} |2\rangle$ but crucially $|2\rangle  \overset{\S}{\longrightarrow} - |1\rangle$. The negative sign simply appears from anticommutation of the fermion fields and implies $\S^2=-1$ acting on the composite bosons leading to the Kramers structure. A similar argument was used to discuss Kramers structure of composite fermions in \cite{Geraedts16} and will appear below establishing the quantum numbers of monopoles. 

\subsubsection{Dual composite boson description}
The single layer conveniently thought of as a chiral (class AIII) TI surface had a dual description in terms of neutral composite fermions, that are $4\pi$-vortices with Kramers doublet structure under particle-hole symmetry, $\S^2=-1$. These composite fermions form the surface state of a helical (class AII) topological insulator coupled to an emergent $U(1)$ gauge field. Can we obtain a similar dual composite particle picture of the half filled bilayer?

For the single layer, the dual description was obtained by examining the surface consequences of bulk electromagnetic duality in the $N=1$ chiral (class AIII) topological insulator. Let us employ the same tactic for the bilayer. We start by by gauging the $U(1)$ symmetry in the bulk of the $N=2$ $3d$ chiral TI. The symmetry and topological properties of the resulting $U(1)$ gauge theory were derived in \cite{Wang16U1SL}, and here we briefly review these results. The key question to address is: what are the symmetry and statistics of the neutral monopole, which will be the bulk prototype for the dual composite particle. 

To compute the properties of monopoles in the $N=2$ state, it is useful to start in the limit of zero charge-coupling so that the surface consists of two free Dirac cones. Next, introducing a strength-one monopole in the bulk 
induces two fermionic zero modes, $\Psi_{1,2}$, 
(one for each Dirac node), which can each be either occupied or empty, corresponding to $\pm 1/2$ charge respectively. Denote the possible states of the vacuum plus monopole configuration with zero-mode occupation numbers $n_{1,2}=\{0,1\}$ by: $|n_1,n_2\> = \(\Psi_1^\dagger\)^{n_1}\(\Psi_2^\dagger\)^{n_2}|0,0\>$. These states respectively have electric charge $q_e=\sum_{i=1,2}(n_i-1/2)\in\{0,\pm 1\}$. The neutral, $q_e=0$, sector contains two states, $|1,0\> = \Psi_1^\dagger|0,0\>$ and $|0,1\> = \Psi_2^\dagger|0,0\>$. Since $\S:\Psi_i^{\vphantom\dagger}\rightarrow \Psi_i^\dagger$, these two neutral monopole states are interchanged by $\S$, and therefore are forced by symmetry to have the same energy. Moreover, these two states transform under $\S$ like a Kramers doublet: $\S\begin{pmatrix}|1,0\> \\ |0,1\> \end{pmatrix} = \begin{pmatrix}|0,1\> \\ -|1,0\> \end{pmatrix}$ (to see this, one can simply choose a gauge in which $\S:|0,0\>\rightarrow |1,1\>$, and $\S:\Psi_i\rightarrow \Psi_i^\dagger$, the result being independent of gauge choice). 

All of the $|n_1,n_2\>$ states are bosons. This can be seen by starting with a unit magnetic monopole in a vacuum with $\theta=0$, and turning up $\theta$ to $2\pi$ to obtain the AIII TI (this can be done while preserving the bulk gap by breaking $\S$ at intermediate stages where $0<\theta<2\pi$, which will not affect the monopole statistics). Turning on $\theta=2\pi$ produces a magnetoelectric polarization giving the monopole unit electric charge, resulting in a dyon with electric and magnetic charge $(q_e,q_m)=(1,1)$ corresponding to the state $|1,1\>$ described above. Since the polarization charge cannot alter the exchange statistics of this particle, the $(1,1)$ dyon remains a boson. Moreover, binding an electron to screen away its charge and obtain a neutral $(0,1)$ monopole does not change its bosonic statistics, since the fermion statistics of the added charge is cancelled by the mutual statistics between the electron and the (1,1) dyon.


Assembling these observations, we see that the neutral monopoles are bosonic Kramers doublets under $\S$ symmetry, having $\S^2=-1$. Like the $N=1$ case, $\S$ acts on the magnetic charges like time-reversal rather than particle-hole symmetry (i.e. is anitunitary and does not flip the magnetic charge). Thus, the $(0,1)$ bosonic monopoles have symmetry group $U(1)_m\rtimes\S$, i.e. are in class AII. Moreover, if we view the $(0,1)$ bosonic monopole as the ``charge" of the theory, its dual ``monopole", with $(q_e,q_m)=(1,0)$, is a fermion (the original electron). The fermionic nature of $(1,0)$ implies that the $(0,1)$ boson monopoles are in a non-trivial SPT phase protected by $U(1)_m\rtimes\S$ symmetry. 
Indeed, the statistical transmutation of a monopole from fermion to boson, dubbed the ``statistical Witten effect", is the hallmark of a bosonic topological insulator phase protected by $U(1)_m\rtimes \S$, which has dual $\theta$-angle  $2\pi$.\cite{Vishwanath13,Metlitski13} 

Then, repeating the arguments of \cite{Metlitski15}, from the bulk electromagnetic duality, and the bulk-boundary correspondence, we see that the $N=2$ AIII surface state, has a dual description in terms of neutral, bosonic $2\pi$-vortices that are Kramers doublets under $\S$, couple to an emergent electromagnetic gauge field, and form the anomalous surface state of a bosonic TI with bulk $\theta=2\pi$. Comparing to the $N=1$ case, we see that apart from the obvious difference in statistics, the composite bosons of the $N=2$ surface carry half of the flux compared to composite fermions for the $N=1$ case. 

Hence, the half-filled bilayer can be described by the dual composite boson effective action:
\begin{align}
\mathcal{L}_\text{dual} = \mathcal{L}_\text{bTI}[a]+\frac{adA}{2\pi}+\frac{AdA}{4\pi}
\label{eq:Lbilayer}
\end{align}
where $\mathcal{L}_\text{bTI}[a]$ describes the anomalous surface state of the $3d$ boson TI with bulk $\theta=2\pi$, coupled to the emergent gauge field, $a$. The flux, $\frac{\nabla\times\v{a}}{2\pi}$ corresponds to the electron density, and the second term captures the minimal coupling of the electron current and the external gauge field, $A$. Finally, the last term appears due to the unconventional non-local action of $\S$ symmetry, under which $\mathcal{L}\rightarrow \mathcal{L}'+2\frac{AdA}{4\pi}$, where $\mathcal{L}'$ represents the ordinary transformation by the local part of $\S$, and the extra Hall conductance term appears due to the non-local action of exchanging full and empty Landau levels. In the following sections, we will analyze various possible boson TI surface states for $\mathcal{L}_\text{bTI}$, and analyze the corresponding states of the half-filled bilayer.

\subsubsection{Quantum Hall ``ferromagnet"}
Since the composite bosons are placed at finite density due to the applied magnetic field, the conceptually simplest possibility is that they can condense to form a superfluid. Any such condensate of unit gauge charge breaks the $\S$ symmetry, due to the Kramers structure of the composite bosons. Additionally, the condensate will produce a Higgs mass for the emergent gauge field $a$, resulting in a trivially confined phase with no fractional excitations. In the half-filled bilayer context, this condensate does not introduce any extra physical Hall conductance, and hence results in a trivial insulator with Hall conductance $\sigma^{xy}=1$.

In terms of the original electrons, this state corresponds simply to a quantum Hall ``ferromagnet", i.e. an inter-layer coherent state in which the electrons fill half of the orbitals of the lowest LLs to form a trivial Hall insulator with broken $\S$. Such an interlayer coherent Hall insulator state may happen spontaneously due to repulsive inter-particle interactions that lead to effective attraction in the exchange channel, favoring spontaneous orbital polarization -- a quantum Hall ``ferromagnet" in which half of the available orbitals are spontaneously filled, and the other half remain empty, resulting in a non-fractionalized quantum Hall insulator $\sigma^{xy}=1$.\cite{Nomura06}

The remaining piece of the puzzle to confirm this correspondence, is to understand how the gapped electron excitations of the quantum Hall ferromagnet appear in the dual composite boson description.  To see this, we will closely parallel the analysis of \cite{Vishwanath13} for gauge-neutral bosons at the surface of a helical bosonic SPT, but incorporating the effects of a fluctuating $U(1)$ gauge field relevant to the composite boson system dual to the half-filled bilayer. Since the composite boson fields, $b_a$, form Kramers doublets under $\S$ symmetry, they have at minimum two components, which we will denote by pseudospin labels $a\in\{\up,\down\}$. We will work in the limit of an enlarged symmetry, where these pseudo-spin species are separately conserved, and later (if necessary) address the consequences of breaking this separate spin and charge conservation down to just a single $U(1)$. The charge-1 superconductor corresponds to a condensate of $\<b_{a}\>=\rho_0e^{i\theta_{a}}$. They key property which arises due to the system being a helical SPT surface state, and which distinguishes it from that of a conventional superfluid, is that the vortex in $b_{\up}$ condensate carries half a unit of gauge charge of the $\down$ species.\cite{Vishwanath13} There are, thus, four species of vortex excitations $\psi_{\pm,a}$, two for each spin species $a$, which carry $\pm1/2$ spin-charge of the opposite spin-boson.  As in any dual description of a superfluid, the vortex fields are non-local objects, and have long range interactions that can be viewed as being mediated by gauge fields $\alpha_{\up,\down}$, whose magnetic flux corresponds to the boson densities: $j_{b_a}^\mu = \frac{1}{2\pi}d\alpha_a$.  Following \cite{Vishwanath13}, we can write an effective field theory in terms of either pair of vortex fields: $\psi_{\pm, \up}$ or $\psi_{\pm, \down}$; here we choose to work with $\psi_{\pm, \down}$:

%
%

\begin{align}
\mathcal{L}_\text{bTI-SF} = \sum_{s=\pm }\frac{1}{2}|\(i\d_\mu-\alpha_{\down,\mu}-\frac{s}{2}a_\mu\)\psi_{s,\down}|^2+\frac{\alpha_\down da}{2\pi}
\label{eq:Lv}
\end{align}
In this description, the vortices $\psi_{\pm,\down}$ can be viewed as fractions of $b_\up = \psi^\dagger_{+,\down}\psi_{-,\down}$, and the density of $b_\down$ is the given by $\frac{1}{2\pi}\nabla\times\vec{\alpha}_\down$. 

This easy-plane $CP^1$ theory is self-dual\cite{Senthil04,geraedts2013exact}, and the $\S$ symmetry, which exchanges $b_{\up}$ and $b_\down$, acts as a duality transformation. To make this more precise, let us assume that the easy-plane $CP^1$ theory in Eq.~\eqref{eq:Lv} does have a continuous critical point. A naive anti-unitary particle-hole transform $\S_0$ acts in Eq.~\ref{eq:Lv} as
\bea
(a_t,a_{x,y})&\to&(a_t,-a_{x,y}), \nonumber \\
(\alpha_t,\alpha_{x,y})&\to&(\alpha_t,-\alpha_{x,y}), \nonumber \\
\psi_{\pm}&\to&\psi_{\pm}.
\eea
This transformation leaves everything invariant except for the mutual Chern-Simons term $\alpha da$, which acquires a minus sign under $\S_0$. To recover the sign of this Chern-Simons term, we perform a self-duality transform by transforming $\psi_{\pm}$ to their vortices. After a few lines of algebra by integrating out trivial degrees of freedom, we obtain a Lagrangian that takes exactly the same form as Eq.~\eqref{eq:Lv}. Thus the particle-hole symmetry $\S$ should be implemented as $\S_0$ followed by a self-duality (particle-vortex) transform. In particular, since the boson mass term $m_+|\psi_{+, \downarrow}|^2+m_-|\psi_{-, \downarrow}|^2$ is invariant under $\S_0$ but changes sign under the self-duality transform, it is odd under the combined particle-hole transform $\S$. This implies that the simple condensate $\<b_{\downarrow}\>\neq0$, corresponding to positive mass for both $\psi_{\pm,\downarrow}$, breaks $\S$ symmetry -- consistent with the fact that $\S^2=-1$ on $b$. In Appendix~\ref{app:dual} we provide another derivation of the duality between the two-flavor Dirac fermion theory and the bosonic gauge theory Eq.~\eqref{eq:Lv}, including the nontrivial implementation of $\mathcal{S}$ symmetry on the bosonic side.

In the superfluid phase, both of the vortex species $\psi_{\pm,\down}$ are gapped. 
The physics of this state can be seen more clearly if we rewrite the gauge fields $a_{\pm}=\alpha_\down\pm a/2$. The above Lagrangian then becomes (after ignoring gapped $\psi$ fields at low energy):
\begin{align}
\mathcal{L}=\frac{1}{4\pi}(a_+da_+-a_-da_-)+\frac{1}{2\pi}(a_+ - a_-) dA+\frac{1}{4\pi}AdA,
\end{align}
where the last two terms come from incorporating the boson TI surface state, $\mathcal{L}_\text{bTI-SF}$ into the full action, Eq.~\ref{eq:Lbilayer} for the half-filled bilayer. The above theory clearly describes a non-fractionalized state, with Hall conductance $\sigma^{xy}=1$, and gapped fermion excitations that are $2\pi$ flux of $a_\pm$.


Hence, we see that the superconductor of $b$ is indeed the $\S$ breaking quantum Hall ferromagnet. The low-energy collective excitations of the resulting quantum Hall ferromagnet depend on the underlying symmetries of the bilayer. The composite boson description above applies for the case in which there is no separate conservation of charge in each layer. In this case, the quantum Hall ferromagnet was fully gapped. However, in the absence of interlayer tunneling, this interlayer charge difference is conserved, and the system can be described as two independent copies of the composite Dirac liquid. Here, the quantum Hall ferromagnet may arise via spontaneous interlayer exciton condensation, resulting in a fluctuating ``superfluid" mode\cite{Fertig89,Yang96}, with  measurable consequences such as interlayer exciton supercurrent drag effects.\cite{Kellogg03} 

\subsubsection{Symmetry preserving anomalous fractional quantum Hall state}
\label{sec:N2TO}
A more interesting possibility arises if the exchange channel interactions are frustrated, e.g. leading to strong quantum fluctuations that can disorder the quantum Hall ferromagnet leading to a $\S$ preserving state. Since this system corresponds to the surface of a $3d$ SPT, the resulting $\S$ preserving state can only be fully gapped if there is an accompanying topological order. The dual description in terms of gauge-charged bosons that form the surface state of a (gauged) helical boson SPT in class AII (symmetry $U(1)_g\rtimes\S$ ), is useful for identifying possible such gapped, symmetric quantum disordered liquids. 

To explore this possibility, we first ignore the fluctuations of the emergent gauge field, $a$. Then, the composite boson system is identical to the surface of a bosonic TI with bulk electromagnetic $\theta$-angle $2\pi$\cite{Vishwanath13} .This boson TI surface can exhibit a quantum disordered symmetry preserving state with $\Z_2$ topological order, whose emergent quasiparticles include bosonic spinons $(e)$ and visons $(m)$ both carrying half-unit of the composite boson charge. This state can be obtained by phase disordering the composite boson superconductor described in the previous section, by condensing two-fold vortices $\<\psi_{+\up}\psi_{-\up}\>=\<\psi_{+\down}\psi_{-\down}\>\neq 0$ such that $\S$ symmetry is restored\cite{Vishwanath13}. The $e$ and $m$ particles are then remnants of the single unpaired vortices. These particles see each other as $\pi$ fluxes, i.e. acquire a phase of $\theta_{e,m} = (-1)$ upon braiding. A notable difference to the cases studied in \cite{Vishwanath13}, is that due to the Kramers structure of the dual bosons in the present case, $\S$ symmetry interchanges the $e$ and $m$ particles.

Upon incorporating fluctuations of $a$, this phase is no longer gapped due to the gapless photon mode $a$ and, in fact, corresponds to an electron pair superfluid state of chiral TI surface. To obtain a fully gapped phase without breaking $\S$ symmetry, we may pair condense the composite bosons, with $\<b_\up^2\> =\<b_\down^2\>\neq 0$. This pair condensation has three important effects. First, a Higgs mass is produced for $a$, so that the state is now fully gapped and topologically ordered. Second, $\pi$-vortices $v$ of the composite boson pair condensate emerge as a new gapped excitations. As for an ordinary superconductor of electrons,  vortices are bound to $\pi$-flux of $a$, and hence have physical electrical charge of $1/2$.
Third, the condensate perfectly screens the emergent gauge charge of $a$, such that formerly $a$-charged excitations are bound to an appropriate fraction of the condensed boson pair to become charge neutral. Due to this screening a neutralized particle formerly with $a$-charge $q$ obtains statistical phase of $e^{i\pi q}$ upon encircling a vortex.

Denoting the ``neutralized" particle obtained from screening away the $a$-charge of a particle $X$ as $\tilde{X}$, the resulting topologically ordered state has excitations: $\tilde{e},\tilde{m},\tilde{\eps}=\tilde{e}\times\tilde{m},\tilde{b},v^n,\dots$ and various combinations. The $\tilde{e},\tilde{m},\tilde{\eps}$ particles have the same self- and mutual statistics as their un-neutralized counterparts. However, $\tilde{e},\tilde{m}$ have mutual statistics $e^{i\pi/2}$ with the vortex $v$, since they are screened by a fraction of the condensed boson pair. The resulting phase has the topological structure of a $\Z_4$ gauge theory in which $\tilde{e}$ and $v$ are respectively the unit $\Z_4$ ``charge" and ``flux" excitations respectively. The theory also contains a fermion (the electron): $c=\tilde{b}\times\tilde{\eps}\times v^2$, that has trivial statistics with all other particles, and physical electrical charge-1 (coming from the $2\pi$ flux of $a$ bound to the $v^2$ part). The anyon types, symmetry properties, and exchange statistics of this anomalous $\Z_4$ topological order are summarized in Tables~\ref{tab:symmetry}~and~\ref{tab:statistics}.

Precisely the same surface topological order can be obtained directly in the fermion language, by phase disordering the superconducting two Dirac cone surface of the $N=2$ class AIII TI.\cite{Wang14,Metlitski14} Here, we have obtained the same result using the dual boson description to illustrate the equivalence of these two perspectives.

In the context of the half-filled bilayer quantum Hall system, this topologically ordered phase will have unit Hall conductance, as for the $\S$-breaking quantum Hall ferromagnet described above. Despite having the same Hall conductance, this topologically ordered phase is distinct from the $\S$-breaking quantum Hall ferromagnet in that it 1) preserves symmetry, and 2) has fractionalized anyonic excitations that carry half-charge and transform as Kramers doublets under $\S$ symmetry. Experimentally detecting these fractional charges may be challenging since there are likely no additional edge modes associated with the $\Z_4$ order since $\S$ symmetry is broken at the boundary, thus eliminating the possibility of detecting these fractional charges through shot-noise or edge tunneling measurements. A simpler, but less direct test, would be that, at half-filling in each layer and in the absence of interlayer tunneling, the $\S$-breaking interlayer coherent state would exhibit a Goldstone mode resulting from interlayer exciton condensation, whereas, the topological ordered state would be fully gapped. Such differences could be measured in local compressibility measurements, or by looking for the presence or absence of exciton-supercurrent drag effects when separately contacting the two layers. Actually, the above $\Z_4$ topological order  preserves the $U(1)_1\times U(1)_2$ symmetry associated with particle-number conservation in each of layers $1$ and $2$ (see Appendix \ref{app:U12}). If interlayer tunneling is also absent at the edge then the $\Z_4$ topological order will possess a non-trivial edge structure, which is responsible for the separate layer Hall-conductances $\sigma^{(1)}_{xy} = \sigma^{(2)}_{xy} = \frac12$ being quantized (see Appendix \ref{app:U12}). These edge states could be potentially probed experimentally.

\renewcommand{\arraystretch}{1.3}
\begin{table}
\begin{tabular}{C{0.9in}C{0.3in}C{0.6in}C{0.6in}C{0.6in}}
\toprule
 & $\theta$ & charge & $\S$-partner & $\S^2$ \\ 
\midrule
$\tilde{e}$ & $1$ & $0$ & $\tilde{m}$ &  \\
$\tilde{m}$ & $1$ & $0$ & $\tilde{e}$ &  \\
$\tilde{\eps} = \tilde{e}\times\tilde{m}$ & $-1$ & $0$ & $\tilde{\eps}$ & $-1$  \\
$b$ & $1$ & $0$ & $\tilde{b}$ & $-1$ \\
$v^n$ & $1$ & $\frac12$ & $v^{-n}$ &  \\
$c = b\times\tilde{\eps}\times v^2$ & $-1$ & $1$ & $c^\dagger$ & \\
\bottomrule
\end{tabular}
\caption{{\bf Symmetry properties -- } for symmetry preserving anomalous fractional quantum Hall state of the half-filled bilayer. Here, $b$ is the dual composite boson which forms a $\Z_2$ topological order, and then pair condenses. $\tilde{e}$, $\tilde{m}$, and $\tilde{\eps}$ are respectively the charge, flux, and fermion of the $\Z_2$ topological order. $v^n$ is an $n$-fold vortex of the dual boson pair superconductor, and $c$ is the local electron. $\theta$ denotes topological spin of the particles (see Table~\ref{tab:statistics} for mutual statistics). Heading ``charge" denotes physical electron charge, ``$\S$-partner" denotes the anyon type that a given anyon transforms to under particle-hole symmetry $\S$, and $\S^2$ is only well-defined only for anyon types that are invariant under $\S$.}
\label{tab:symmetry}
\end{table}

\renewcommand{\arraystretch}{1.3}
\begin{table}
\begin{tabular}{C{0.3in}|C{0.3in}C{0.3in}C{0.3in}C{0.3in}}
 & $\tilde{e}$ & $\tilde{m}$ & $b$ & $v^n$\\ 
\hline
$\tilde{e}$ & $1$ & $-1$ & $1$ & $i^n$ \\
$\tilde{m}$ & $-1$ & $1$ & $1$ & $i^n$ \\
$b$ & $1$ & $1$ & $1$ & $(-1)^n$ \\
$v^n$ & $i^n$ & $i^n$ & $(-1)^n$ & $1$ 
\end{tabular}
\caption{{\bf Topological properties -- } for symmetry preserving anomalous fractional quantum Hall state of the half-filled bilayer (see Table~\ref{tab:symmetry} for definitions and symmetry properties).}
\label{tab:statistics}
\end{table}

\subsection{Half filled quadruple layer $(N=4)$}
\label{sec:N4}
With these examples in hand, we may freely continue to generalize to even higher Chern numbers. Physically relevant examples with Chern bands of $N=4$ include spin-polarized bilayer graphene in the zeroth LL limit, and $4$-layer GaAs systems at half-filling in each layer. Generalizing the previous two examples, the half filled $N=4$ case corresponds to a four-Dirac cone surface state of a $3d$ chiral TI. In the presence of interactions, this surface state is topologically equivalent to a bosonic SPT surface state, protected by $\S$-reversal alone -- i.e. with no anomalous action of charge conservation symmetry \cite{Fidkowski13,Wang14,Metlitski14}. Closely related to this, the bulk magnetic monopole in this $N=4$ chiral TI phase has trivial $\S$ quantum numbers, implying that the phase does not permit a dual description in terms of point-like dual composite ``vortices". Nevertheless, the perspective of the higher dimensional SPT surface is useful for identifying possible phases of the half-filled quadruple layer.

The simplest candidate state for this 4-Dirac cone SPT surface is a spontaneously $\S$ breaking quantum Hall ferromagnet with $\sigma^{xy}=2$. Another possible fate of the quantum disordered Hall ferromagnet, previously known from the chiral TI surface perspective\cite{Fidkowski13,Wang13Window,Wang14,Metlitski14}, is that the system can form a fractionalized phase that preserves $\S$ symmetry, at the expense of developing $\Z_2$ Abelian topological order. In this anomalous topological order the elementary $\Z_2$ charge, $e$, and flux, $m$, excitations have integer charge but anyonic mutual statistics, and transform as Kramers doublets under the non-local $\S$ symmetry. Directly confirming this topological order by measuring the presence of anyonic excitations can in principle be done using interferometric techniques, but may be experimentally challenging. It remains an open question of how one may more realistically detect the anomalous symmetry properties of the excitations in this $\Z_2$ topological order.
A third intriguing possibility is that the system may develop some larger emergent symmetry, which protects a fluctuating gapless phase. We will examine this prospect below.

\subsection{Half-filled octuple layer $(N=8)$}
Finally, we conclude the discussion of Chern bands by examining the case of Chern number eight, i.e. of the half-filled octuple-layer quantum Hall system. In the absence of interactions, this system corresponds to the surface of an chiral TI with eight surface Dirac cones. While this eight Dirac cone state has non-trivial band topology, in the presence of interactions, this $3d$ TI phase becomes topologically equivalent to a trivial insulator.\cite{Fidkowski13,Wang14,Metlitski14} Hence, it is possible for interactions to fully gap out the half-filled octuple-layer quantum Hall state to result in an insulator with Hall conductance $\sigma^{xy}=4$, which fully preserves the non-local $\S$ symmetry. This corresponds to a quantum-disordered quantum Hall ferromagnet, which for $N=8$, need not exhibit any topologically non-trivial excitations. 

While the resulting state is a featureless paramagnet, this result is conceptually interesting, as it is enabled purely by strong correlation effects. Namely, without interactions, it is impossible to make a $\S$-preserving band-insulator out of the lowest Landau level orbital of the octuple-layer quantum Hall system at half-filling. Moreover, despite the collapse of the $3d$ TI invariant under interactions for $N=8$, the corresponding $2d$ Chern bands remain non-trivial and distinct topological bands regardless of interactions effects. Hence, we find a simple example where a half-filled topological band in $2d$ corresponds to a trivial insulator in higher dimensions.

We note that both the featureless paramagnet in this section and the $\Z_2$ topological order in section \ref{sec:N4} break  particle number conservation in each layer, and, as a result, only preserves $\S$ which acts identically on all layers. In fact, as long as particle-number conservation in each layer is preserved, $\S$ symmetry implies that the Hall conductivity of each layer is $\sigma_{xy} = 1/2$, so no featureless state is allowed for any number of layers $N$. For any $N$, topological order that preserves particle number conservation in each layer can  be constructed, e.g. by stacking $N/2$ copies of $N  = 2$ $\Z_4$ order in appendix \ref{app:U12}. This topological order will support non-trivial edge-states necessary to maintain $\sigma_{xy} = \frac12$ in each layer.

\section{Multicomponent landau levels with larger symmetries}
In this section, we consider half-filled multicomponent Landau level systems with additional local symmetries. We start with Abelian symmetries, such as a conserved spin-component ($U(1)_\text{spin}$). Next we explore how additional non-Abelian symmetries can lead to protected gapless states, previously discussed in the context of deconfined criticality.\cite{Senthil04,Nahum15} Examples include SU(2) orbital symmetry in a two-component, or SO(5) symmetry in four-component LL systems.

\subsection{$N=2$ with $U(1)_\text{spin}$}
Above, we have considered the TI surface correspondence of half-filled multicomponent Landau levels with only charge conservation and non-local particle-hole symmetry. In the following sections, we examine the SPT surface state correspondence for $2d$ systems with additional continuous symmetries. One such possibility is having a half-filled quantum spin-Hall state, in which there is a conserved component of spin resulting in an additional $U(1)_\text{spin}$ symmetry, and in which opposite spin-species have opposite filling fractions, e.g.: $\nu_\up = +1/2$, and $\nu_\down = -1/2$. Such a quantum spin Hall state at integer filling, has been achieved experimentally in twisted bilayer graphene\cite{Young14}, and its half filled analog would correspond to the surface of a $3d$ electronic SPT phase with $U(1)_c\times U(1)_\text{spin}\rtimes \S$, which can be viewed as  two copies of $N = 1$ AIII TI, with opposite magnetic fields applied to the two copies. Each copy has a dual description via the composite fermion theory (\ref{eq:Lcf}), so we get
\be L = \sum_{s = \up/\down} \left(\bar{\psi}_{\text{cf},s}(i\slashed{\partial}+{\slashed{a}}_s-\mu\gamma_0)\psi_{\text{cf}_s} -\frac{1}{4\pi} a_s dA_s\right)\ee
where $A_{\up/\down}$ is the external gauge field coupling to  $\up/\down$  electrons. Note that that the two Dirac cones have opposite doping $\psi^{\dagger}_{\text{cf},\up} \psi_{\text{cf},\up} = -\psi^{\dagger}_{\text{cf},\down} \psi_{\text{cf}, \down}$. We note also, that breaking the $U(1)_\text{spin}$ state by allowing interlayer tunnelling or pairing, in a way that preserves $\S$-symmetry, will lead to the dual composite boson description discussed above in the absence of spin-conservation. The composite bosons have $2\pi$ flux, and can be viewed effectively as fractions of the $4\pi$ flux object $\psi_{\text{cf},\up} \psi_{\text{cf},\down}$ in the spin conserving case.


\subsection{$N=2$ with $SU(2)$}
For half-filled two component LL systems, such as a bilayer quantum well of GaAs, or single-layer graphene at neutrality, the maximal possible symmetry is a full $SU(2)$ symmetry rotating between identical orbitals in each layer. Such a symmetry is approximately present in single-layer graphene, in which the long-range part of the Coulomb interactions do not distinguish between the different sub-lattices, and hence the physics with this enlarged $SU(2)$ symmetry may provide a reasonable description of single-layer graphene at intermediate energy scales.

The presence of $SU(2)$ symmetry is known\cite{Wang14} to prevent the possibility of having a gapped symmetry preserving surface state in the corresponding two-Dirac cone SPT surface states of bilayer GaAs or single layer graphene (which has two valleys).  Given the exact correspondence between the SPT surface states and the $2d$ systems with non-local symmetry, this symmetry enforced gaplessness also extends to the half filled two-component systems with SU(2) symmetry. Hence, the resulting phase must either be gapless correlated phase or break symmetry.

\subsection{$N=4$ with SO(5) symmetry}
We conclude the discussion of higher symmetries by exploring the possibility of higher symmetries that can emerge in four-component LL systems. For example, the long-range part of Coulomb interactions in graphene and bilayer graphene does not distinguish between layer or valley, and exhibits a large $SU(4)$ symmetry rotating among various quantum Hall ferromagnetic orders. This symmetry is broken at short-distances by lattice scale effects, but in certain regimes, e.g. achieved by tuning the in-plane and out-of-plane components of magnetic field to the phase boundary between antiferromagnetic and Kekule disorted phases, a large SO(5) subgroup of the full flavor symmetry may remain intact~\cite{Senthil04,LeeSachdev14,LeeSachdev15,Wu14,Nahum15}. In this section we explore how, when combined with the non-local $\S$ symmetry, this enlarged SO(5) symmetry can protect gapless critical behavior related to that discussed in deconfined critical points between antiferromagnetic and valence-bond solid orders~\cite{Senthil04}. We note that, since the appearance of this paper, a more detailed study of symmetries and dualities for deconfined critical points appeared in Ref.~\onlinecite{wang2017deconfined}.



We first consider the surface of the corresponding $U(1)\times\S$ topological insulator with four Dirac cones, with surface Hamiltonian:
\begin{align}
H_\text{surface} = \sum_{a=1}^4 \psi_a^\dagger\(p_x\sigma^x+p_y\sigma^z\)\psi_a^{\vphantom\dagger}
\end{align}
where $\S: \psi_a \rightarrow i\sigma^y\psi_a^\dagger$. 

Upon adding a magnetic field to this surface and projecting into the zeroth Landau levels at half-filling, the kinetic energy of the electrons is quenched. We begin by considering non-interacting mass terms that would gap out the zeroth Landau level, of the form $\psi^\dagger_a M_{ab}\sigma^y\psi^{\vphantom\dagger}_b$, with $M_{ab}$ a Hermitian matrix. There are 16 total such mass terms $M_{ab}$, which decompose into an SO(5) singlet, $M_{ab}=\delta_{ab}$, a 5-component SO(5) vector (fundamental representation): $M_{ab}\in \Gamma^{\alpha}=\{\vec{\tau}\eta^z,\eta^{x,y}\}$ where, $\alpha=1,2,\dots 5$, and $\vec{\tau},\vec{\eta}$ are Pauli matrices, and 10-component adjoint representations, $\Sigma^{\alpha\beta}=i\[\Gamma^\alpha,\Gamma^\beta\]$. The SO(5) symmetry excludes the latter $15$ fundamental and adjoint terms as these transform nontrivially under symmetry.  On the other hand, the singlet mass, $\psi^\dagger\sigma^y\psi^{\vphantom\dagger}$, which would produce a trivial quantum Hall insulator, is not forbidden by the SO(5) symmetry, and is only ruled out by $\S$ symmetry. 

Since bare mass terms are symmetry forbidden, the zeroth Landau level is highly degenerate at the single-particle level, and interactions inherently dominate. A conceptually simple outcome that has been explored extensively theoretically\cite{Nandkishore10,Jung11} and experimentally\cite{Feldman09,Zhao10} is that the exchange effects drive a spontaneous ordering into a quantum Hall magnet, by spontaneously forming a mass term. Without the SO(5) symmetry, the microscopic interactions can explicitly pick out a preferred ferromagnetic ordering direction out of the set of masses. However, the enhanced SO(5) symmetry rules out explicit anisotropies, e.g. of the form $-\sum_\alpha V_\alpha\(\psi^\dagger\Gamma^\alpha \sigma^y\psi^{\vphantom\dagger}\)^2$ with different strengths, $V_\alpha$, for different mass directions, and allows the quantum Hall magnetic ordering direction to freely fluctuate within the SO(5) vector manifold.

To describe the resulting fluctuating quantum Hall ferromagnet, we can introduce non-linear sigma model (NL$\sigma$M) for the fluctuating quantum Hall ferromagnet. 
Namely, we can decompose a generic density-density interaction term by introducing a five-component bosonic field $n^\alpha$:
\begin{align}
\mathcal{L}&_\text{int} = \int_{r,r'}(\sum_a\psi_a^\dagger\psi_a^{\vphantom\dagger})_rV(r-r')(\sum_a\psi_a^\dagger\psi_a^{\vphantom\dagger})_{r'}
\nonumber \\
&\rightarrow \frac{1}{2\lambda}(\partial_\mu n^\alpha)^2-r_\alpha|n^\alpha|^2+u_{\alpha,\beta}|n^\alpha|^2|n^\beta|^2+\sum_\alpha n^\alpha \psi^\dagger \Gamma^\alpha\psi
\label{eq:InteractionHS}
\end{align}
The SO(5) symmetry requires $r_\alpha$ to be independent of $\alpha$ and $u_{\alpha,\beta} = u$.
The quantum Hall ferromagnet phase is described by a state in which one of the components of $n^\alpha$ develops a non-zero expectation value.  Alternatively, if quantum fluctuations in the direction of $n$ are strong, spontaneous symmetry breaking may be avoided in lieu of a gapless quantum fluctuating state that preserves symmetry.  Fixing the magnitude of $n$, but allowing its orientation to fluctuate, the fermionic modes will be gapped everywhere, and can be integrated out to yield the following topological non-linear sigma model\cite{Abanov00,LeeSachdev14,LeeSachdev15}:
\begin{align}
\mathcal{L}_\text{NL$\sigma$M} &= \frac{1}{2\tilde{\lambda}}(\partial_\mu n^\alpha)^2+2\pi i \Gamma_\text{WZW}[n]
\nonumber\\
\Gamma_\text{WZW}[n] &= \frac{1}{4!\Omega_4}\int_\mathcal{B}d^4x \epsilon^{\mu\nu\lambda\rho}\epsilon^{\alpha\beta\gamma\delta\epsilon} n_\alpha\d_\mu n_\beta\d_\nu n_\gamma\d_\lambda n_\gamma \d_\rho n_\epsilon \label{eq:WZW}
\end{align}
plus the constraint $\sum_{\alpha=1}^5 n_\alpha^2 =1$. Here $\Omega_4 = \frac{2\pi^{5/2}}{\Gamma(5/2)} = \frac{8\pi^2}{15}$ is the surface area of a 4-sphere, and $\mathcal{B}$ is the Ball obtained by extending the physical space-time, $X$, to $X\times[0,1]$, where $n(x_{1,2,3},x_4)$ interpolates between some fixed reference configuration at $x_4=0$, to the physical configuration at $x_4=1$.  The presence of the topological term was previously derived by \cite{LeeSachdev14,LeeSachdev15}. In the present context, the appearance of this topological term is natural given the correspondence with the nontrivial SPT surface state.

Thus we see that the combination of SO(5) and $\S$ symmetries can protect a critical state with gapless bosonic excitations. This state arose previously in studies of deconfined phase transitions between antiferromagnetic and valence-bond solid states.\cite{Senthil04} The crucial difference with the present system is that  strictly in 2d with a local anti-unitary symmetry $\S$, one cannot have the theory (\ref{eq:WZW}) emerge with $\S$ acting as $\S: n^{\alpha} \to -n^{\alpha}$. 
However, the theory (\ref{eq:WZW}) with such a symmetry action may be realized in bilayer graphene\cite{Wu14} in the quantum limit with a non-local $\S$ symmetry.

\section{Half Filled Topological Insulator Flat Bands}
\subsection{Two Dimensional Flat bands and $3d$ Topological Phases in Class CII }
We begin with by establishing that a microscopic model of a local Hamiltonian that exhibits a topological insulator flat band with a non-local particle hole symmetry. We intend this discussion as a proof of principle for the existence of such topological flat bands in local Hamiltonians, rather than a realistic proposal for a particular material. Alternatively, the scheme we will outline could be used as a numerical device for simulating correlated topological insulator surface states without the extra computational complexity of simulating the accompanying TI bulk.

\subsubsection{Microscopic Model of $2d$ TI Flat Bands }
The key ingredient to obtaining a non-local particle-hole symmetry was having a dispersionless (flat) topological band. In the previous examples based on multicomponent Landau levels, the flatness was readily obtained by applying an appropriate orbital magnetic field. However, in other examples, such as the time-reversal symmetric topological insulator without any spin-conservation, there is no such vector potential that can achieve the desired flatness. However, even in these cases, one may still formally construct local models with exactly flat topological bands, via a different route\cite{Neupert11}, which we will now describe. This construction can be directly implemented numerically, to simulate higher dimensional SPT surface states, and may also approximately describe physical materials in which interactions are much stronger than the single-particle bandwidth.

Consider a non-interacting tight-binding Hamiltonian $H_0= \sum_{k}c_{ka}^\dagger\mathcal{H}_{ab}(k)c^{\vphantom\dagger}_{kb}$, where $c_{ka}$ annihilates an electron with momentum $k$ and spin/orbital index $a\in\{1\dots N\}$, and $\mathcal{H}(k)$ is an $N\times N$ matrix with eigen-energies $\varepsilon_n(k)$ labeled by band index $n$, and corresponding wave-functions $\phi^n_{ka}$. Suppose the band associated with eigenvalues $n=1\dots r$ has a particular non-trivial (symmetry protected) topological invariant, $\nu$, and is separated by an energy gap from those with $n=r+1\dots N$. 

As a concrete example, let us consider the Kane-Mele model\cite{kane2005quantum} of the $2d$ helical topological insulator (class AII), which consists of a time-reversed pair of Haldane's honeycomb models of a Chern insulator\cite{haldane1988model}, with opposite Chern number for opposite spin species, plus a Rashba spin orbit coupling term that preserves $\T$ while removing any unnecessary spin-conservation symmetry. This model has two orbitals, $i=1,2$, and two spin species $\sigma=\up,\down$, such that the indices $a=(i,\sigma)$ can take four values. The band structure has a two separate pairs of bands, which we will label $n=1,2$ and $n=3,4$. $\e_1(k)-\e_2(k)$ (and similarly $\e_3(k)-\e_4(k)$) vanish at time-reversal invariant momenta ($k=-k+G$ for some reciprocal lattice vector, $G$), but the gap between the pairs remains non-zero everywhere, $\e_{34}(k)-\e_{12}(k)>\Delta$. The orbitals of bands $1,2$, together, exhibit a non-trivial $\Z_2$ topological invariant (similarly for those of $3,4$), and we may consider flattening this topological band.

Then, we may construct a flattened Hamiltonian\cite{Neupert11}, formed from projection operators into the orbital basis spanned by the topological band:
\begin{align}
H_F^0 &= E_0\int \frac{d^dk}{(2\pi)^d} c^\dagger_{ka}\sum_{n=1}^r\Pi_{k,ab}^{n}c^{\vphantom\dagger}_{kb}
\nonumber\\
\Pi^n_{k,ab}&\equiv \delta_{ab}-\phi^n_{ka}\phi^{n*}_{kb}
\end{align}

The Hamiltonian $H_F$ has the following properties: $H_F$ contains a dispersionless zero energy band with non-trivial topological invariant $\nu$, preserves the protecting symmetry, and is local (i.e. contains couplings that fall off exponentially in distance)\cite{Neupert11}. On the face of it, one might suspect that the absence of a local (symmetry preserving) Wannier basis for the single particle wave-functions $\phi^n_{ka}$ would prevent either the preservation of symmetry or the locality of $H_F$. However, the absence of a symmetry preserving Wannier basis stems from an obstruction to forming a globally smooth phase for the wave-functions $\phi^n_{ka}$. In contrast, the phase of the wave-functions drop out of the projection operators $\Pi^{(n)}(\v{k})$, so that these projectors have a smooth, analytic k-space structure, and hence exponentially decaying real-space structure.

The Hamiltonian, $H_F^0$ is invariant under the anti-unitary particle-hole operation:
\begin{align}
\S: |0\>\rightarrow \prod_{k}\prod_{n=1}^r\psi_{nk}^\dagger|0\>
\nonumber\\
\psi_{nk} \equiv \sum_a \phi^n_{ka}c_{ka}
\end{align}
where $|0\>$ is the state with no electrons. Physically, the band-projection is a reasonable description when the inter-band gap $E_0$ greatly exceeds the interaction strength $V$. 

\subsubsection{Symmetry allowed interaction terms}
We next ask: which interaction terms can be added to the flat-band Hamiltonian while preserving the non-local particle-hole symmetry? In the half-filled Landau level case, which was defined in a system with continuous translation invariance, any two-body interaction term, upon lowest LL projection, could be made to preserve $\S$ simply by tuning the uniform chemical potential to maintain half-filling. In the present case, due to the essential role of the lattice, only a subset of two-body interactions will be suitable. To identify potential $\S$-preserving interaction terms, we first consider $\S$-transformation properties of a generic local fermion bilinear, $\rho_{ab}(r) = c_{ra}^\dagger c^{\vphantom\dagger}_{rb}$, after projection, $\Pi$, into the topological flat band:
\begin{align}
\rho^{\Pi}_{ab} &\equiv \Pi \rho_{ab}(r)\Pi 
\nonumber\\
&=\sum_{n,m=1}^r\sum_{k,k'}e^{i(k-k')\cdot r}\phi^n_{ka}\phi^{m*}_{k'b}\psi_{kn}^\dagger\psi_{k'm}^{\vphantom\dagger}
\end{align}
Under the anti-unitary particle-hole transformation, $\S$, this bilinear transforms as:
\begin{align}
\S\rho_{ab}^\Pi \S^{-1} = \bar\rho_{ab}- \rho_{ba}^{\Pi}
\end{align}
where:
\begin{align}
\bar{\rho}_{ab} = \sum_{n=1}^r\sum_k \phi^{n*}_{ka}\phi^n_{bk}
\end{align}
is the average value of the bilinear within the orbitals of the topological band. 

A general interaction term of the form $H_V=\frac{1}{2}\sum_{r,r'}\rho_{ab}^\Pi(r)V_{ab,cd}(r-r')\rho^\Pi_{cd}(r')$, then transforms as:
\begin{align}
\S H_V\S^{-1} = H_V - \sum_{r,r'}\bar\rho_{ab} V^{ab,cd}(r-r')\rho^{\Pi}_{cd}(r')+\text{const}
\end{align}
i.e. is generically \emph{not} invariant under $\S$ due to the second term. 

The $\S$-breaking effects of a given interaction $V$ can be removed by considering the modified interaction, $\rho_{ab}^\Pi\rightarrow \rho_{ab}^\Pi-\frac{1}{2}\bar\rho_{ab}$ in $H_V$, though, for the most general interactions such counter-terms will be very complicated and fine-tuned. However, for certain simple classes of interactions such as site-local density-density interactions, and in the presence of additional lattice symmetries, the counterterms required to preserve $\S$-invariance amount merely to a simple uniform shift of the chemical potential. 

To see a concrete example, let us again consider the Kane-Mele honeycomb model for the $2d$ AII TI described above, and restrict our attention to site-local density-density interactions, for which the interaction $V$ decomposes as $V_{ij,kl}=V_{ik}\delta_{ij}\delta_{kl}$. For this model, there is a $C_2$-rotation symmetry that exchanges the two sub-lattice sites of the honeycomb: $i=1\overset{C_2}{\longleftrightarrow} 2$. Then, the counterterm $\sum_{r,r'}\frac{1}{2}\bar\rho_{ij}V_{ij,kl}(r-r')\rho^\Pi_{kl}(r') = \delta \mu \sum_i\rho^\Pi_{ii}$, just takes the form of an overall shift in the chemical potential. 

To summarize, in this model, we may add any such density-density interaction while preserving the non-local $\S$ symmetry, simply by renormalizing the chemical potential to maintain half-filling. Therefore, while the $\S$-symmetry is not as robust as in the translationally invariant half-filled LL case, for which translation symmetry automatically fixes all two-body interaction terms to be $\S$-invariant, there is nevertheless a broad class of interacting flat-band models with half-filled non-trivial AII bands which exhibit a non-local $\S$ symmetry, without fine tuning.

\subsection{Surface State of $3d$ Class CII Topological Phase}
Having shown that the topological insulator bands can be flattened and driven into a strongly interacting regime with non-local particle-hole symmetry, we now explore this system from the perspective of the higher dimensional surface state.
For weak interactions, the surface state of the $3d$ class CII topological insulator, can be described by a pair of 2+1D Dirac fermions:
\begin{align}
H_\text{CII} = v\psi^\dagger \v{p}\cdot {\boldsymbol\sigma} \tau^z \psi
\end{align}
where $\sigma$ and $\tau$ are independent sets of Pauli matrices, and the symmetries act like: $\T\psi = i\sigma^y K \psi$, and $\C\psi = i\sigma^y\tau^x\psi^\dagger$. Crucially, the charge-conjugation symmetry squares to minus one, $\C^2\psi = (-1)\psi$, which plays an essential role in ruling out the $\T$-symmetric mass term $m \psi^\dagger \tau^x\psi$. Adding this $\T$-invariant, but $\C$-breaking mass term gaps out the surface states, but with opposite $2d$ AII invariants for $m>0$ and $m<0$. Since $m$ changes sign under $\C$, we see that this surface state is in a sense, symmetrically tuned to the critical point between the $2d$ helical TI and trivial phases. Note further, that this is natural for a surface state that corresponds to a half-filled $2d$ AII TI band -- which is also poised equally between TI and trivial.

\subsubsection{Dual composite boson description}
While we have started with  a non-interacting description, since the $2d$ trivial and topological insulators remain distinct phases even in the presence of arbitrary symmetry preserving interactions, the intermediate critical point and hence the corresponding $3d$ surface state will also be stable to interactions as well, i.e. it cannot be gapped into a trivial insulating phase without breaking symmetry. Proceeding in a similar manner as above for other two-Dirac cone TI surface states, we may construct a dual description of this CII surface state, but examining the bulk electromagnetic duality via the surface-bulk correspondence. The symmetry properties of the bulk monopoles are identified in Appendix~A. We note that in contrast to a previous study\cite{Wang14}, we find that the neutral bulk magnetic monopole transforms has $\C^2=-1$, and trivial $\S^2$. This bosonic monopole has a corresponding surface excitation -- a $2\pi$-flux bosonic vortex, which couples to an emergent $U(1)$ gauge field and serves as the basis for a dual composite boson description of the CII Dirac surface state. The distinct anomalous feature of the composite boson is its transformation property under charge conjugation, $\C^2=-1$.

While the dual description of the half-filled time-reversal invariant topological insulator flat bands bears a close resemblance to that for the half filled quantum Hall bilayer, there is a key physical distinction. Namely, whereas dual composite bosons in the half-filled quantum Hall bilayer occur at finite density, those in the half-filled TI flat-band occur at zero density. This can be seen simply by symmetry considerations: the composite boson density is odd under time-reversal symmetry, and hence must be zero in a time-reversal invariant system. The energetics of the resulting systems may then be rather different. Namely, while finite density bosons will naturally condense to form a trivial particle-hole breaking quantum Hall ferromagnet, those at zero density may more easily form more exotic insulating states. Thus the resulting physics of these two half-filled flat bands may naturally form insulating states, that are necessarily exotic if they preserve symmetry. We leave a detailed study of the energetics for realistic interactions for future work, and instead explore which anomalous topologically ordered fractional states are in principle possible.

\subsubsection{Anomalous topological order}
As for previously discussed examples of two Dirac cone surface states with dual composite boson descriptions, the CII surface state (and hence also the half filled $2d$ TI band) may be driven by strong correlations into an anomalous $\Z_4$ topological ordered state, which preserves the symmetries, but cannot be realized in a purely local $2d$ system. The detailed properties of this phase are derived in Appendix~A from the point of view of phase-disordering the CII surface paired superfluid by condensing four-fold vortices. The key  feature of the resulting $\Z_4$ topological order, which makes it anomalous, is that the neutral bosonic generator of $\Z_4$ differs by a neutral fermion compared to its $\C$-conjugation partner.

\subsection{Half-filled topological insulator in $3d$ - Anomalous Boundary state of a 4+1D topological phase}
Interestingly, the above construction identifying a correspondence between the $2d$ helical TI flat band and the $3d$ CII surface state works directly in one higher dimension. Namely, a half-filled $3d$ helical TI (symmetry $U(1)\rtimes \T$, class AII) flat-band exhibits a non-local particle-hole symmetry, $\C$, and realizes the surface physics of a $4d$ TI with symmetry $U(1)\rtimes(\C \times\T)$ (class CII). This $4d$ TI again has a $\Z_2$ band-invariant, and has surface states that may be detached from the bulk and flattened in the same manner as described above. This correspondence provides an interesting example where a non-local symmetry enables access to surface states of a topological insulator in an inaccessibly high spatial dimension. 

The non-interacting $3d$ surface state of the $4d$ TI (class CII) has one (four-component) Dirac fermion:
\begin{equation}
\label{4dDiracCII}
H_\text{$4d$S}=v\psi^{\dagger}\v{p}\cdot\boldsymbol{\sigma}\tau^z\psi,
\end{equation}
where $\sigma$ and $\tau$ are Pauli matrices. Time-reversal acts as $\mathcal{T}: \psi\to i\sigma^y\psi$ and the unitary charge conjugation acts as $\mathcal{C}: \psi\to i\sigma^y\tau^x\psi^{\dagger}$, with $\T^2=\C^2=(-1)^F$. A $\C$-breaking mass term $m\psi^{\dagger}\tau^x\psi$ leads to either a trivial or topological insulator (class AII), depending on the sign of the mass parameter $m$. This theory can therefore be viewed as a critical point between a trivial and a topological insulator. Notice the equations take essentially the same form as the $2d$ surface state of the $3d$ TI (class CII), except the Hamiltonian contains an additional term $\sim p_z\sigma^z\tau^z$.

Notice that, unlike its lower dimensional counterpart, here, time-reversal symmetry is not crucial for protecting the Dirac fermions in Eq.~\eqref{4dDiracCII}. In fact, the $3d$ surface state is protected (anomalous) as long as the $U(1)\rtimes \mathcal{C}$ symmetry is preserved. The anomalous nature of this state can be understood as an easy-plane anisotropic version of the $SU(2)$ Witten anomaly\cite{wittenanomaly}. A simple way to understand the Witten anomaly with $U(1)\rtimes \mathbb{Z}_2^C$ symmetry is as follows: suppose we add a $\mathcal{C}$-breaking mass term
\be
\label{CIImass}
H_\text{mass}=m\psi^{\dagger}(\tau^y{\rm sin}\theta+\tau^x{\rm cos}\theta)\psi,
\ee
where $m$ and $\theta$ are parameters. Under $\mathcal{C}$ we have $\theta\to \theta+\pi$. It is also known that the $3d$ surface electro-magnetic $\Theta$-angle, defined through the topological response term
\be
\mathcal{L}_{\Theta}=\frac{\Theta}{8\pi^2}dAdA,
\ee
is given by $\Theta=\theta+\Theta_0$ where $\Theta_0$ is a constant independent of $\theta$. It follows that the nonlocal (anomalous) $\mathcal{C}$ symmetry shifts $\Theta\to\Theta+\pi$. Similar relation holds for the gravitational $\Theta_g$-angle.

It is then obvious why such a theory is anomalous: if there were a trivial gapped surface state preserving the $U(1)\rtimes \mathcal{C}$ symmetry, we would require $\Theta=\Theta+\pi ({\rm mod}\,  2\pi)$ (and likewise for $\Theta_g$), which is impossible to satisfy. The free Dirac fermion evades this problem by being gapless, hence having no well-defined $\Theta$-angle.

On the other hand, if the system develops an intrinsic topological order, with fractionalized excitations carrying fractional charge, the effective period of the $\Theta$-angle can be reduced\cite{fracTI1,fracTI2}. Likewise if there exists fractionalized excitations that are charge-neutral fermions, the effective period of the gravitational $\Theta_g$-angle will also reduce to $\pi$\cite{gravtheta}, enabling a gapped $\C$-invariant fractionalized surface state. This implies that it may be possible to make the surface Dirac fermions gapped, by having intrinsic topological orders. We will construct such a gapped state below.

\subsection{A $3d$ topological order for half-filled TI}
To construct a gapped $\C$ symmetric topological surface state for the $4d$ CII topological insulator, and hence also its $3d$ flat band counterpart, we begin with a parton construction in which we fractionalize the electron operator into $n$ charge $e/n$ bosons and a neutral fermion:
\be
c=b^nf,
\ee
This description contains a $u(1)$ gauge redundancy generated by: $b\to e^{-i\alpha}b$, $f\to e^{in\alpha}f$, which is captured in the low energy effective theory as an emergent $3d$ gauge field, $a$. To reproduce the free fermion theory when the $b$ are condensed into a superfluid, we will take an ansatz in which the $f$ fermions form the surface Dirac fermion as in Eq.~\eqref{4dDiracCII}. The low energy theory is then described by
\be
\label{CIIparton}
\mathcal{L}^{(n)}[\psi,b,a_{\mu}]=\bar{\psi}i\slashed{D}_{na+A}\psi+\mathcal{L}_{b}[b,a_{\mu}],
\ee
where $a$ is the dynamical gauge field generated from the parton decomposition, $A$ is a background gauge field coupling to the physical electromagnetic $U(1)$ symmetry current, and $D_{na+A}$ represents covariant derivative with gauge field $na+A$. If we further condense $\<b\>\neq0$, then $c\sim f$ and we recover the free Dirac theory in Eq.~\eqref{4dDiracCII}. But if we keep $b$ gapped, we have a distinct surface state.

A crucial point is that the theory in Eq.~\eqref{CIIparton} is not necessarily  well defined for all $n$. To see this, we will adopt a similar approach to that in~\cite{seiberg2016gapped}. Intuitively, because $a_{\mu}$ is an emergent gauge field that lives purely within the $3d$ surface, it cannot be responsible for any anomalous properties extending from the bulk. 

To test which fractionalization patterns result in a non-anomalous $\C$-invariant action for $a$, let us re-run the argument below Eq.~\eqref{CIImass} within the parton description. Suppose we give the Dirac fermions a mass as in Eq.~\eqref{CIImass}. Then the $\mathcal{C}$-operation, which inverts the mass, generates an extra term in the gauge field Lagrangian:
\bea
\Delta\mathcal{L}&=&\frac{\pi}{8\pi^2}d(na+A)d(na+A)  \nonumber \\
&=&\frac{\pi}{8\pi^2}dAdA+\frac{n\pi}{4\pi^2}dadA+\frac{n^2\pi}{8\pi^2}dada
\eea
The first term, as discussed above, simply represents the anomalous response to an external electromagnetic field. The other two terms involve $a_{\mu}$, and therefore should be trivial (i.e. have the same charge/monopole lattice is a $U(1)$ gauge theory in the trivial vacuum) if Eq.~\eqref{CIIparton} is well defined. For the last term $dada$ to be trivial, we need $n$ to be even, else there is a topological magnetoelectric effect that produces half electric charge dyons. For the second term $dadA$ to be trivial, we need $n=4k$. The factor of four is somewhat subtle, since all dyons have integer charges for $n=2$. However, the dyon statistics become transmuted from those of the conventional $U(1)$ gauge theory in a trivial vacuum due to a Witten effect\cite{wittendyon} in which a monopole of $a_{\mu}$ acquires unit charge under $A_{\mu}$. This charge can then be neutralized by attaching an electron, but at the expense of switching the statistics of the monopole from bosonic to fermionic. This statistical transmutation shows that that the $dadA$ term is anomalous for $n=2$, but is absent for $n=4k$. We note, in passing, that a formal way to summarize this argument is that $a_{\mu}$ is an ordinary $U(1)$ gauge field, while $A_{\mu}$ is actually a $spin_c$ connection.

Therefore the minimal parton decomposition in Eq.~\eqref{CIIparton} that allows for $\C$-symmetry is that with $n=4$. To get a gapped phase, we can simply pair-condense the Dirac fermions by having
\begin{equation}
\mathcal{L}_\text{pairing}=i\Delta\psi^T\sigma^y\tau^z\psi+h.c.,
\end{equation}
with $\<\Delta\>\neq0$. Both the Dirac fermions $\psi$ and the gauge field $a_{\mu}$ acquires a gap from this pair condensate, so the entire system is in a gapped phase. The $\mathcal{C}$ and $\mathcal{T}$ symmetries are explicitly preserved. Therefore this construction gives a gapped symmetric phase. Since the condensate $\psi\psi$ carries gauge charge $q_a=8$, the $U(1)$ gauge symmetry is broken to $\mathbb{Z}_8$: the final state is a $\mathbb{Z}_8$ gauge theory.

The fundamental gauge charge in the $\mathbb{Z}_8$ gauge theory is represented by the $b$ boson. The fermion pair condensate sets $a_{\mu}= -A_{\mu}/4$ at long wavelengths. Therefore the $b$ boson, originally coupled minimally to $a_{\mu}$ before the pair-condensate, now couples minimally to $A_{\mu}$ as charge-$1/4$ object. This simply means that the gauge charge $b$ carries physical charge $1/4$. 

Let us show how the existence of the charge-$1/4$ excitation reduces the periodicity of $\Theta$ such that $\Theta$ becomes equivalent to $\Theta+\pi$, so that the gapped surface theory is compatible with $\C$-symmetry. It suffices to show that $\Theta=\pi$ is trivial in the presence of the fractional bosons. For this, we examine the monopoles of $A_{\mu}$. The strength-$1$ monopole ($q_m=1$) acquires charge $q_e=\Theta/2\pi=1/2$, which can be canceled by attaching a $b^2$ boson. Crucially, due to the existence of the charge-$1/4$ boson $b$, the $q_m=1$ monopole cannot exists on its own because it violates the Dirac quantization rule. It can exist, however, as the end point of a $\mathbb{Z}_8$ gauge flux line -- the violation of Dirac quantization then simply comes from the gauge flux in the flux line. Since the flux line has finite tension (energy cost proportional to length), the monopole is confined, and therefore does not have meaningful statistics or gauge charge. Similarly the $q_m=2$ monopole is also confined. The $q_m=4$ monopole, however, is not confined, and it is meaningful to ask about its statistics and gauge charge. Now the bare $q_m=4$ monopole has electric charge $q_m=4\Theta/2\pi=2$. This can be canceled by attaching a physical Cooper pair $cc$. Since the Cooper pair is bosonic and carries no $\mathbb{Z}_8$ gauge charge, the neutralized $q_m=4$ monopole will also be bosonic and gauge-invariant. Therefore $\Theta=\pi$ is indeed trivial in the presence of the $\mathbb{Z}_8$ gauge theory with charge-$1/4$ excitations.



The above argument also suggests that charge fractionalization is required for gapped phases of Witten anomaly with $U(1)\rtimes \mathbb{Z}_2$ symmetry. This means that if the symmetry is enhanced from $U(1)\rtimes\mathbb{Z}_2^C$ to full $SU(2)$ (corresponding to the original Witten anomaly), such topological orders cannot exist, because $SU(2)$ symmetry cannot be fractionalized. The only option for a symmetric state is then to be gapless, like the free Dirac fermions. This is a three-dimensional analog of symmetry-enforced gaplessness first discussed in two dimensions in \cite{Wang14}.

We also notice that gapped phases for Witten anomaly with $U(1)\rtimes\mathbb{Z}_2^C$ symmetry were first discussed in \cite{youxu}, in which a $\mathbb{Z}_2$ topological order was proposed. Our argument suggests that further fractionalization (for example, to $\mathbb{Z}_8$ topological order) seems to be required.

%

\section{Discussion}
Generalizing the previously studied case of the half-filled Landau level, we have uncovered a number of examples of D-dimensional topological flat bands that exhibit a non-local particle hole symmetry when half-filled, enabling them to reproduce the anomalous surface state physics of a (D+1)-dimensional topological insulator surface state in one lower dimension. In cases where the (D+1)-dimensional surface state permits a non-interacting limit with two flavors of Dirac fermions, we identified a dual composite boson description of the half-filled topological band in terms of neutral vortices of the electron fluid. We also identified candidate symmetry preserving fractionalized phases for the half-filled topological flat bands, and explored their potential physical manifestations in multilayer quantum wells, and single- and bi- layer graphene in the quantum Hall regime. Finally, we constructed a general prescription for this type of ``dimensional enhancement" of topological flat bands, and identified a $3d$ example that realizes the anomalous surface physics of a physically inaccessible $4d$ topological insulator. While we have explored several candidate systems based on general symmetry and anomaly grounds, an important future task will be to conduct a detailed study of the energetics of these phases within model Hamiltonians, and analyze whether and how exotic anomalous fractionalized phases may be realized and detected under realistic experimental conditions.

\textbf{Acknowledgments}: We thank Maissam Barkeshli, Parsa Bonderson, Fiona Burnell, and Meng Cheng for helpful conversations. ACP was supported by the Gordon and Betty Moore Foundation EPiQS Initiative through Grant GBMF4307. CW is supported by the Harvard Society of Fellows.  Research at Perimeter Institute for Theoretical Physics (MM) is supported by the Government of Canada through the Department of Innovation, Science and Economic Development and by the Province of Ontario through the Ministry of Research and Innovation.   AV was supported by a Simons Investigator grant and by the ARO MURI on topological insulators, grant ARO W911NF-12-1-0461.

\textbf{Remark}: For complementary work on multicomponent half-filled quantum Hall systems, developed in parallel to ours by Sodemann, Kimchi, Wang, and Senthil, see Ref.~\cite{sodemann2016composite} which appeared simultaneously with this work. 
\appendix 

\section{Alternative derivation of the two-component duality}
\label{app:dual}

We now derive the duality between a theory of two Dirac fermions:
\be
\label{twoflavorDirac}
\mathcal{L}=\sum_{i=1,2}\bar{\psi}_ii\slashed{D}_A\psi_i,
\ee
and a gauge theory with bosonic matter fields $z_{\pm}$:
\be
\label{twoflavorboson}
\mathcal{L}=\sum_{s=\pm}\left(\left|D_{(\alpha+\frac{s}{2}a)}z_{s}\right|^2-|z_{s}|^4\right)+\frac{1}{2\pi}\alpha da+\frac{1}{2\pi}adA,
\ee
where $A$ is the background probe gauge field, $a,\alpha$ are emergent dynamical gauge field, and $D_a$ denotes covariant derivative with gauge connection $a$. The term $|z_s|^4$ is a short hand notation to denote that the $z_s$ bosons are tuned to the Wilson-Fisher fixed point. Following \cite{karch2016particle,seiberg2016duality,Murugan}, we derive this duality starting from a simpler duality known as $(2+1)d$ bosonization. This bosonization duality states that a single Dirac fermion
\be
\mathcal{L}=\bar{\psi}i\slashed{D}_A\psi
\ee
is dual to
\be
\label{bosonize}
\mathcal{L}=\left|D_{\alpha}z\right|^2-|z|^4+\frac{1}{4\pi}\alpha d\alpha+\frac{1}{2\pi}\alpha dA+\frac{1}{8\pi}AdA,
\ee
where $A$ is the probe gauge field, $z$ is a Wilson-Fisher boson and $\alpha$ is an emergent $U(1)$ gauge field. Taking time-reversal conjugate of the above equation, it is easy to see that a single Dirac fermion is also dual to
\be
\label{Tbosonize}
\mathcal{L}=\left|D_{\alpha}z\right|^2-|z|^4-\frac{1}{4\pi}\alpha d\alpha-\frac{1}{2\pi}\alpha dA-\frac{1}{8\pi}AdA.
\ee

\begin{widetext}
It appears that Eq.~\eqref{bosonize} and \eqref{Tbosonize} describe two different theories, but in fact they are simply two dual descriptions of the same theory. To see this, simply perform another boson-vortex duality on the $z$ bosons in Eq.~\eqref{Tbosonize} and we get
\bea
\mathcal{L}&=&|D_{\alpha'}z'|^2-|z'|^4-\frac{1}{2\pi}\alpha d\alpha'-\frac{1}{4\pi}\alpha d\alpha-\frac{1}{2\pi}\alpha dA-\frac{1}{8\pi}AdA \nn
&\to&|D_{\alpha'}z'|^2-|z'|^4+\frac{1}{4\pi}\alpha' d\alpha'+\frac{1}{2\pi}\alpha' dA+\frac{1}{8\pi}AdA,
\eea
where the last line came from integrating out $\alpha$. This action then has the same form as Eq.~\eqref{bosonize}. Therefore the $z$ bosons in Eq.~\eqref{bosonize} and \eqref{Tbosonize} are mutual vortices of each other.

Now back to the two-flavor Dirac fermion Eq.~\eqref{twoflavorDirac}. We use the bosonization duality Eq.~\eqref{bosonize} on $\psi_1$ and the time-reversed form Eq.~\eqref{Tbosonize} on $\psi_2$. Adding together the actions we get
\be
\mathcal{L}=\sum_{i=1,2}\left(\left|D_{\alpha_i}z_{i}\right|^2-|z_{i}|^4\right)+\frac{1}{4\pi}(\alpha_1d\alpha_1-\alpha_2d\alpha_2)+\frac{1}{2\pi}Ad(\alpha_1-\alpha_2).
\ee
Now redefine $z_1=z_+$, $z_2=z_-$, $\alpha_1=\alpha+\frac{1}{2}a$ and $\alpha_2=\alpha-\frac{1}{2}a$, we get exactly the desired dual action in Eq.~\eqref{twoflavorboson}.
\end{widetext}

In the main text we discussed how time-reversal symmetry should act as particle-vortex duality on the $z_{\pm}$ bosons (also known as the self-duality of easy-plane $CP^1$ theory). The above derivation makes this obvious, since for each single component of Dirac fermion, the dual $z$ boson should transform to its vortex under time-reversal.

\section{Half-filled bilayer ($N= 2$) with intra-layer $U(1)$ symmetry.}
\label{app:U12}
In this Appendix, we derive an $S$-preserving topological order for the $N  =2$ bilayer assuming a $U(1)_1\times U(1)_2$ symmetry corresponding to conservation of particle number in each layer (here the subscripts $_1$ and $_2$ are meant to distinguish layers, and do should not be confused with the related notation for a $U(1)$ Chern-Simons theory at level $k$). Actually, we will obtain the same ${\mathbb Z}_4$ topological order as in section \ref{sec:N2TO}; however, now we will be able to determine the quantum numbers of anyons under both $U(1)_1$ and $U(1)_2$. This will, in turn, tell us about the edge structure in the presence of $U(1)_1\times U(1)_2$ symmetry.

We begin by considering the layers as decoupled. In this case, each layer can develop a CT-Pfaffian topological order. In principle, such a (CT-Pfaffian)$^{\otimes 2}$ state is an acceptable $S$-symmetric topological order of the bilayer. However, we can obtain a simpler Abelian symmetry preserving state as follows. 

Recall\cite{Metlitski15STO,Wang13,Metlitski15,Wang15} that CT-Pfaffian can be thought of as a subset of Ising$\times U(1)_{-8}$. The anyons are $1_k$, $\sigma_k$ and $f_k$, where $\{1, \sigma, f\}$ run over Ising charge and $k = 0\ldots 7$ labels the $U(1)_{-8}$ charge. For even $k$ only $1_k$ and $f_k$ are allowed, and for odd $k$ - only $\sigma_k$. The physical electric charge of the anyon is $k/4$. The physical electron $c = f_4$. Under ${\cal S}$, $k \to -k$, so that $1_2 \to f_{-2}$, $f_2 \to 1_{-2}$. $f_0$ is a Kramers doublet under $S$. Now, we have two-copies of CT-Pfaffian: we will label the anyons $\{1^a_k, \sigma^a_k, f^a_k\}$ with  superscript  $a = 1,2$ corresponding to the layer index. We  point out that the anyons $f^1_4$ and $f^2_4$ are simply electrons in corresponding layers, so they are topologically identical (even though they carry different charges under $U(1)_1\times U(1)_2$).

 Now, imagine condensing the bosonic, neutral $S$-singlet anyon $f^1_0 f^2_0$.  Now, anyons which differ by $f^1_0 f^2_0$ are identified, e.g. $f^1_0 \sim f^2_0$  - we denote this simply as $f_0$. Moreover, some non-Abelian anyons split into multiple Abelian particles. For instance, the quantum dimension 2 anyon $\sigma^1_1 \sigma^2_{-1}$ splits into two Abelian anyons $e$ and $e \times f_0$, both of which are bosons and carry charge $q = (1/4, -1/4)$ under $U(1)_1\times U(1)_2$. We have $e \times e = 1^1_2 f^2_{-2}$. Now, label $v  = e \times 1^{2}_2$ - this is also a boson with $q  = (1/4, 1/4)$. The anyons $e$ and $v$ have mutual statistics $+i$. It turns out that the topological order of the condensed phase is just ${\mathbb Z}_4 \times \{1, c\}$ with the electric and magnetic generators of ${\mathbb Z}_4$ being $e$ and $v$. This is the same topological order we found in section \ref{sec:N2TO}, however, now we know the quantum numbers of $e$ and $v$ under the layer $U(1)$ symmetries.
 
What about the action of $S$-symmetry in the condensed phase? There are actually two options which are both possible:
\bea \mathrm{Option\,\,1:}\quad v \to v^{\dagger}, \quad e \to e v^2 c^1\nonumber\\
\mathrm{Option\,\,2:}\quad v \to e^2 v c^1, \quad e \to e^{\dagger}\eea
where the superscript on $c$ denotes the layer to which the electron belongs. Option 1 is the one we found in section \ref{sec:N2TO} (and, in fact, $e v^2 c^1$ corresponds to the anyon $m$ of that section). Option 2 is a different $S$-symmetric topological order of the bilayer.

Now, we can ask about the edge states of the above phases. Since $S$-symmetry will be broken at the edge, the two options above will generally give rise to the same edge spectrum. A possible edge theory can be obtained by starting with the edge of (CT-Pfaffian)$^{\otimes 2}$. Recall, that a CT-Pfaffian supports a right-moving $c = 1$ charged mode (corresponding to the $U(1)_{-8}$ sector )with conductance $1/2$ and a neutral left-moving $c = -1/2$ mode (corresponding to the Ising sector). Now, stacking the two CT-Pfaffians, we have two right-moving $c = 1$  modes, each carrying charge of the corresponding layer and two left-moving neutral $c = -1/2$ modes. After $f^1_0 f^2_0$ condensation, the two-left moving modes can be thought of as one neutral $c  =-1$ mode. Other edge structures are possible, however, as long as there is no tunnelling between the layers at the edge, we cannot have a simple integer-quantum Hall edge, due to the bulk fractional Hall conductances $\sigma^1_{xy} = 1/2$, $\sigma^2_{xy} = 1/2$ for the two layers.

\section{Properties of the $3d$ TI with symmetry $U(1)\rtimes(\T\times\C)$ (class CII)}
In this Appendix, we deduce various properties of the $3d$ class CII TI, whose surface corresponds to the $2d$ time-reversal invariant TI (class AII), in the flat, half-filled limit. We begin by demonstrating that the surface states of this $3d$ TI, in the non-interacting limit, can be detached from the bulk states and flattened, which establishes the precise correspondence between the Hilbert space and Hamiltonian of the flat $3d$ TI surface band and flat lower dimensional band. We then examine the properties of magnetic monopoles in the bulk of the $3d$ CII TI, which, via the bulk-boundary correspondence, yields a dual composite boson description of the surface. Next, we derive the properties of vortices in the surface paired superfluid, which we then use to construct the $\Z_4$ surface topological order.

\subsection{Disconnecting the $3d$ class CII TI surface state from the bulk \label{app:disconnect}}
In the main text, we have noted that $2d$ $\T$-invariant topological insulators (class AII) have a $3d$ counterpart (class CII) with the same band classification, is $\T$-invariant, and also exhibits an extra extra particle-hole symmetry, $\S$. To complete the explicit demonstration that flattening and half filling the $2d$ class AII TI band precisely corresponds to the surface state of this $3d$ class CII TI, we must show that the $3d$ surface state can be disconnected from the bulk bands and deformed into the flat band limit.

To this end, we begin with the conventional two Dirac cone surface state of the  $3d$ class CII TI, which can be described by the Hamiltonian:
\begin{align}
H_\text{CII} = v\psi^\dagger \v{p}\cdot {\boldsymbol\sigma} \tau^z \psi
\label{appeq:HCII}
\end{align}
where $\sigma$ and $\tau$ are independent sets of Pauli matrices, and the symmetries act like: $\T\psi = i\sigma^y K \psi$, and $\C\psi = i\sigma^y\tau^x\psi^\dagger$. Crucially, the charge-conjugation symmetry squares to minus one, $\C^2\psi = (-1)\psi$. We note that adding the $\C$-breaking mass $m\psi^\dagger \tau^1\psi$ gaps out the surface states, but with opposite $2d$ AII invariants for $m>0$ and $m<0$. Since $m$ changes sign under $\C$, then we see that this surface state is poised on the boundary between the $2d$ helical TI and trivial phases -- as expected for the half-filled helical TI flat-band. To demonstrate that this $3d$ CII surface state precisely corresponds to  the TI flat bands, it only remains to show that we may detach the surface state bands from those of the bulk, and deform the surface dispersion to the flat-band limit. 

\begin{figure}[t!]
\includegraphics[width=\columnwidth]{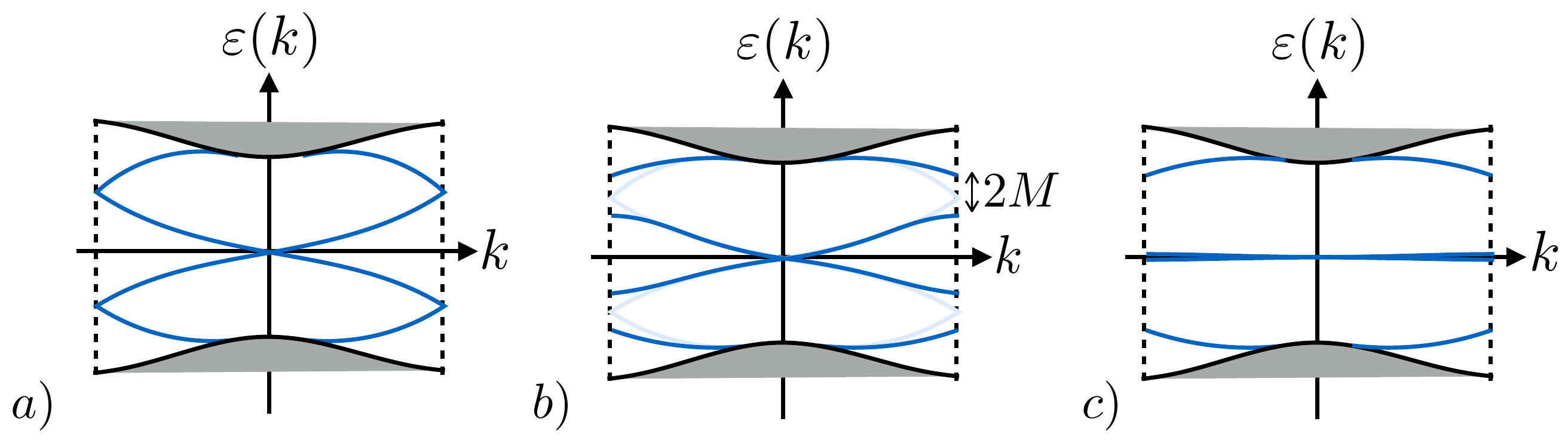}\vspace{-.1in}
\caption{ {\bf Disconnecting the $3d$ Class CII surface and bulk states - } Schematic depiction of the surface projected density of states. Gray shaded regions denote bulk states, blue lines are surface states, and vertical dashed lines mark the surface Brillouin zone boundaries. Starting from a surface termination where the surface states run continuously across the bulk band gap (a), one may add a perturbation to break off a surface band that is completely isolated within the band gap (b), and then consider the narrow bandwidth limit where this isolated band becomes flat, and corresponds to the purely $2d$ flat class AII band (c).}
\label{fig:flattening}
\vspace{-.1in}
\end{figure}

We can detach the surface bands from the bulk using the following steps (see Fig.~\ref{fig:flattening}). First, consider the case that the surface bands meet the time-reversal invariant momenta (TRIM) points at the boundary of the surface Brillouin zone before running into the bulk (this can always be accomplished for sufficiently small velocity $v$). At a fixed boundary TRIM, there is a four-fold intersection of the surface states at energy $E$ and another particle-hole conjugate set of $4$ states at energy $-E$, and we may describe the surface Hamiltonian linearized near these points by $H_\text{TRIM}\approx \psi \(v\delta{\bf k}\cdot {\bs \sigma}\tau^z+E\eta^z\)\psi$, where $\delta{\bf k}$ is the deviation of momenta from the TRIM point, and $\eta^z$ is an additional Pauli matrix that distinguishes the band-indices of the $\pm E$ energy band touchings. We may then add the symmetry-allowed perturbation $M\psi^\dagger\tau^2\eta^2\psi$, which opens a gap about the $\pm E$ points at the Brillouin zone edge, thereby separating the surface bands from those of the bulk. We may then freely deform the surface dispersion to contain a completely flat band lying at zero energy in the center of the bulk gap. The Hilbert space of the flat surface state band hence precisely corresponds precisely to that of that of the $2d$ helical TI band with the same Brillouin zone size.

\subsection{Bulk Monopole}
Consider beginning with the free fermion, double Dirac cone surface of the $3d$ class CII TI surface, Eq.~\ref{appeq:HCII}, and inserting a magnetic monopole source from into the TI bulk from the vacuum. This corresponds to adding a vector potential $A_\lambda(\v{r})$ to the surface with $\oint_S\nabla\times \v{A}\cdot d\hat{n} = 2\pi$. In the presence of such a vector potential, the surface Dirac cone develops two complex fermionic zero modes with definite $\sigma^z = +1$, and $\tau^z=\pm 1$, each of which has half electric charge.

Hence, the bulk monopole created when both of these modes are occupied or empty has electric charge $\mp 1$. Then, to obtain a neutral bulk monopole we must bind it to an electron. Since the electrons of the class CII TI have $\C^2=\T^2=-1$, then the neutral monopole will also have $\C^2=-1$ inside the bulk. Note that $\C^2$ is well defined on both electric and magnetic charges (since it cannot be changed by redefining $\C$ by a U(1) charge rotation of $\C$) whereas $\T^2$ is well defined only on electric charges (since it can by redefined by a magnetic U(1) rotation of $\T$). 

To gain further confidence in the identification $\C^2=(-1)^{n_m}$, where $n_m$ is the monopole number, we first note that, in general, there are precisely two possible projective symmetry actions: the  one proposed above, $\C^2=(-1)^{n_m}$, and the trivial action, $\C^2=1$. Moreover, we can nail down which of these two options occurs by consider a charge-1 dyon in a trivial 3+1D bulk insulator, and adiabatically transforming the bulk into a nontrivial class CII TI along a route that sacrifices $\T$ symmetry but preserving $\C$ symmetry. Along the way, we will continuously turn on a $\theta$ angle that screens away the charge of the dyon, however the $\C^2$ action is sharply quantized and cannot be altered. Hence the resulting (now neutral) monopole retains its initial $\C^2=-1$ property, as predicted above by more naive arguments. To be explicit, let us examine the adiabatic deformation of from trivial to CII TI along a time-reversal breaking path within the following 3+1D bulk model:
\begin{align}
H_\text{bulk} = \psi^\dagger\(\vec{p}\cdot\vec{\sigma}\tau^z\lambda^z+m_1\lambda^x\)\psi^{\vphantom\dagger}
\end{align}
where $\vec{\lambda}$ are an extra set of Pauli matrices describing bulk orbital degrees of freedom. The trivial (vacuum) is described by $m_1>0$, and the class CII TI by $m_1<0$. The symmetry action is as follows: $\T$ acts as above, and $\C:\psi\rightarrow -i\sigma^x\tau^y\lambda^y\psi^\dagger$, which ensures that $H_\text{bulk}$ is $\C$-invariant while maintaining the above described transformation properties for the surface states (which are zero modes of a mass domain wall in $m_1(\vec{r}) = -m_0\text{sign}(z)$). To continuously tune the bulk between TI and trivial without closing the bulk gap, we may add an additional $\C$-preserving, but $\T$ breaking mass: $m_2\psi^\dagger\tau^z\psi^{\vphantom\dagger}$. We can view the resulting state as a pair of ordinary time-reversal protected TIs (class AII), one each for $\tau^z=\pm$, which are protected from mixing by an additional $\C$ symmetry. Starting with a trivial insulator ($m_1=m>0$) we may continuously deform into a CII TI ($m_1=-m<0$) by breaking time reversal with $m_2$, and dialing the $\theta = \tan^{-1}(m_2/m_1)$ angle for these two AII TIs from $0$ to $\pi$, to while keeping fixed non-zero dynamical mass: $\sqrt{m_1^2+m_2^2}=m\neq 0$. As these two $\theta$ angles are dialed from $0$ to $\pi$ any magnetic charge $q_m$ induces an electric polarization charge $\Delta q_e = 2\times\(\frac{\theta}{2\pi}\) q_m\rightarrow q_m$, completing the demonstration of the above argument.

This non-trivial $\C^2=-1$ value of the bulk monopole prevents one from trivially confining the bulk $U(1)$ gauge theory along with its fermionic matter by condensing this monopole, unless one breaks $\C$ or $\T$ symmetry. Moreover, the surface excitation created by this bulk monopole will, like the monopole, be a charge-neutral bosonic with $\C^2=-1$, and is just the composite boson object with $2\pi$ vorticity of the electrons that form the building block of the dual description of the two Dirac cone surface state of the class CII TI.

The extra electron does not change the combined effect of $\C$ and $\T$ symmetry, since $\S^2\equiv \C^2\T^2=(+1)$ for the fermions (note that while $\S^2$ is ill-defined separate for the electron and the charge-one dyon, their combination, which is charge neutral, has well-defined $\S^2$, which may be computed from the separate $\S$ properties of electron and dyon in any convenient fixed gauge). Hence, the bulk neutral monopole of the CII TI has $\S^2=+1$. This result contrast that of a previous analysis of \cite{Wang14}, which predicted $\S^2=-1$. However, the $\S^2=-1$ bulk monopole is actually the defining of a different class of $3d$ TIs -- chiral TIs in class AIII. In particular, the symmetries of class CII constrain the bulk band structure to have a trivial AIII invariant, such that the CII TI can be viewed as two opposite copies of AIII, protected from mixing by $\C$-symmetry.

\subsection{Surface vortex theory}
A useful step towards deriving the properties of a symmetry preserving topologically ordered surface phase is to examine the properties of vortices in the $U(1)$-symmetry breaking surface paired superfluid. We will then obtain the surface topological order by phase disordering the surface-superfluid to restore the $U(1)$ symmetry, by condensing the minimal vorticity object that does not break symmetry.

Beginning with the free fermion description of the $3d$ CII TI surface, we may add a superconducting pairing term: $\Delta  \sum_{a,b=\pm} \psi_{\up,a}\tau^z_{ab}\psi_{\down,b}+h.c.$, which preserves $\C$, $\T$, and $\S$. Enlarging our fermionic fields into an 8-component Nambu spinor: $\Psi \equiv \begin{pmatrix} \psi^{\vphantom\dagger} \\ \T \psi^\dagger \end{pmatrix}$, and introducing Pauli matrices $\eta$ that operate in the Nambu subspace, the surface superfluid Hamiltonian can be written as:
\begin{align}
H_\text{SSF} = \int_{\v{r}} \Psi^\dagger \(v\v{p}\cdot\bs{\sigma}\tau^z\eta^z+\Delta(\v{r})\tau^z\eta^++\Delta^*(\v{r})\tau^z\eta^- \)\Psi
\end{align}

Next consider the effect of a $hc/2e = \pi$ superfluid vortex, e.g. $\Delta(\v{r}) = f(r)e^{i\phi}$, where $\phi = \tan^{-1}(y/x)$ and $f(r)$ is some function interpolating between $f(0)=0$, and $\lim_{r\gg \xi}f(r)=\Delta_0$, where $\xi$ is the vortex core size. The vortex has a pair of Majorana zero modes respectively labeled by $\tau^z=\pm 1$ eigenvalues:
\begin{align}
\gamma_\pm = \int_re^{-\int_0^r f(r) dr}\frac{1}{2}\(\varphi^{\pm 1} \psi_{\down,\pm} (\v{r})+\varphi^{\mp 1}\psi_{\down,\pm}^\dagger(\v{r})\)
\end{align}
where $\varphi = e^{i\pi/4}$. These modes are exponentially localized to the vortex core. Thus each vortex has two degenerate states, $|v,\pm\>$ labelled by fermion parity of the zero modes $i\gamma_+\gamma_-=\pm 1$.

While the vortex configuration breaks $\C$ and $\T$, it preserves their combination $\S$. Under $\S$, the vortex core fermion states transform as $\S:\gamma_\pm \rightarrow \gamma_{\mp}$, so that the mass term $i\gamma_+\gamma_-$ is symmetry allowed, and will generically be present, splitting the energies of the occupied and unoccupied vortex state. Thus there are two types of $\pi$ vortices in the superfluid, which differ by a neutral fermionic quasiparticle. We denote these two vortex configurations as $e_1$ and $m_1$ respectively, where the subscript $1$ refers to number of fluxes in units of $\pi$, and the $e$ and $m$ labels are chosen since these two objects are bosons, but differ by a neutral Bogolyubov quasiparticle and hence have mutual $\pi$ statistics as for the electric and magnetic particles in a $\Z_2$ gauge theory. Similarly, we will denote the two types of $-\pi$ vortices as $e_{-1}$ and $m_{-1}$. Note that, since the $e$ and $m$ vortices have opposite fermion parity of their core fermion modes, $\C$, symmetry will exchange $\C:e_{1}\leftrightarrow m_{-1}$, and $\C:m_1\leftrightarrow e_{-1}$. Moreover, since $\S$ preserves $e_1$ and $m_1$, $\T$ also interchanges $\T: e_{\pm 1}\leftrightarrow m_{\mp 1}$. Therefore, a given vortex, when fused with its $\C$ partner, leaves behind a neutral Bogolyubov quasiparticle.


\subsection{Symmetry preserving surface topological order}
Having worked out the properties of vortices in the surface superfluid, we may now readily construct the surface topological order by condensing the minimal vorticity excitation which has trivial symmetry properties. 
There are no suitable $\pi$ vortex excitations that can be condensed while preserving symmetry. For example consider condensing $e_1$. Since $m_{-1}$ has mutual semionic statistics with $e_1$ (due to the difference in occupation of the core fermion mode), it is impossible to condense both $e_1$ and its $\C$ conjugated partner, $m_{-1}$, and hence the $e_1$ condensate breaks $\C$ symmetry. Identical reasoning rules out symmetry respecting condensates of the other $\pi$ vortices. This obstacle is reassuring, since condensing $\pi$ vortices would result in a trivial insulating phase with no fractionalized excitations, which at an SPT surface must require breaking the protecting symmetry.

What about $2\pi$ vortices? Bound states of two $e_1$ or two $m_1$ particles are topologically trivial bosonic objects, which we will collective denote $I_2$ ($I$ denoting the topological trivial, or ``identity" sector). However, these objects transform non-trivial under charge conjugation, having: $\C^2=-1$, as can be seen by the fact that inserting a neutral monopole into the bulk creates an $I_2$ particle in the surface superfluid, and since the bulk monopole has $\C^2=-1$ (see above), whereas the monopole in vacuum had trivial $\C^2$, the surface $I_2$ vortex must also have $\C^2=-1$.

However, we see that the $4\pi$ vortices $I_4$, made of two $I_2$ vortices, are bosonic and transform trivially under all symmetries, and hence may be condensed (in equal measure with their time-reversed partners $I_{-4}$. The result is a $\Z_4$ topological order, with a neutral fermion $f$ (the remnant of the Bogolyubov quasiparticles of the superfluid), the remnants of the $\pi$ vortices, $\{e_{\pm1},m_{\pm 1}\}$, and a charge-1/2 boson that is $1/4$ of the Cooper pairs of the superfluid and becomes a sharp deconfined quasiparticle in the quadruple-vortex condensate state. The $f$ particle has mutual semionic statistics with the $\pi$ fluxes, and $b$ and $e_\pm,m_\pm$ have mutual statistics $\theta_{b,e_\pm}=\theta_{b,m_\pm} = e^{\pm i\pi/4}$.

Though the $\C^2$ value of the $I_2$ vortex cannot be altered by a gauge transformation, assigning symmetry properties to an object that transforms to a topological sector $I_2\rightarrow I_{-2}$ may make one uneasy. In this case, one can be reassured that an identical conclusion can be reached using the formalism of Barkeshli et al.  \cite{barkeshli2014symmetry}.

Under symmetry the $f$ particle transforms just as the original electron, having $\C^2=\T^2=-1$. The half charge boson, $b$, transforms conventionally under symmetry ($\T$ acts trivially, $\C$ flips its charge and $\C^2=1$). The most notable symmetry action, which captures the anomalous symmetry structure of the SPT surface, is that $\C,\T:e_{\pm 1}\leftrightarrow m_{\mp 1}$.

\bibliography{HalfFilledBib}

\end{document}